\documentclass[useAMS,usenatbib,usegraphicx]{mn2e}

\title[Merger History Predictions from Semi-Analytical Models]{A Comparison of Galaxy Merger History Observations and Predictions from Semi-Analytic Models}   

\author[Bertone \& Conselice]{Serena Bertone$^{1}$\thanks{E-mail: serena@scipp.ucsc.edu},
Christopher J. Conselice$^{2}$ \\
$^{1}$Santa Cruz Institute for Particle Physics, University of California, 1156 High Street, Santa Cruz, Santa Cruz, CA 95064, USA \\
$^{2}$University of Nottingham, School of Physics \& Astronomy, Nottingham, NG7 2RD UK }

\def\solm{M$_{\odot}\,$}

\def\kms{km s$^{-1}$}

\def\rate{$\Re (z)\,$}
\def\hm{$h^{-1}$}
\def\1{$^{-1}$}

\begin{document}

\date{Accepted ; Received ; in original form}

\pagerange{\pageref{firstpage}--\pageref{lastpage}} \pubyear{2009}

\maketitle

\label{firstpage}

\begin{abstract}

Mergers are predicted in all cosmological models involving Cold Dark Matter to be one of the dominant channels whereby galaxies accrete mass. In this paper we present a detailed analysis of predicted galaxy-galaxy merger fractions and rates in the Millennium simulation and compare these with the most up to date observations of the same quantities up to $z \sim 3$. We carry out our analysis by considering the predicted merger history in the Millennium simulation within a given time interval, as a function of stellar mass. This method, as opposed to pair fraction counts, considers mergers that have already taken place, and allows a more direct comparison with the observed rates and fractions measured with the concentration--asymmetry--clumpiness (CAS) method. We examine the evolution of the predicted merger fraction and rate in the Millennium simulation for galaxies with stellar masses $M_{\star} \sim 10^{9} - 10^{12}$ \solm.
We find that the predicted merger rates and fractions match the observations well for galaxies with $M_{\star} > 10^{11}$ \solm\ at $z<2$, while significant discrepancies occur at lower stellar masses, and at $z>2$ for $M_{\star} > 10^{11}$ \solm\ systems. At $z>2$ the simulations underpredict the observed merger fractions by a factor of 4-10.
The shape of the predicted merger fraction and rate evolutions are similar to the observations up to $z \sim 2$, and peak at $1<z<2$ in almost all mass bins. The exception is the merger rate of galaxies with $M_{\star} > 10^{11}$ \solm, which remains high at $z < 1.5$. 
We discuss possible reasons for these discrepancies, and compare different realisations of the Millennium simulation to understand the effect of varying the physical implementation of feedback.
We conclude that the comparison is potentially affected by a number of issues, including uncertainties in interpreting the observations and simulations in terms of the assumed merger mass ratios and merger time-scales. The differences between the observations and simulation results might also be due to problems in the modelling of star formation in the simulation, which produces redder and less biased galaxies than observed, particularly for galaxies with stellar masses $M_{\star} < 10^{11}$ \solm\.
Finally, our findings may also be related to other CDM problems, including the the lack of massive galaxies with $M_{\star} > 10^{11}$ \solm\ at $z > 1$, and a lack of merger events between lower mass galaxies.

\end{abstract}


\section{Introduction}
\label{intro}

Theoretically, the dominant and largely successful paradigm for 
understanding and modelling the universe is based on the theory of Cold Dark 
Matter (CDM).
In this scenario, the first objects that form in the universe are low mass 
haloes that subsequently merge together to form more massive structures as 
time progresses (e.g., \citealt{white1978}). The hierarchical mass assembly 
by merging is in fact the cornerstone of CDM-based simulations and can be 
empirically tested, for example, by investigating the role of mergers in the 
formation and evolution of galaxies (e.g. \citealt{kauffmann1993}; 
\citealt{berrier2006}; \citealt{wetzel2008a}).

Galaxy formation and evolution is certainly driven in part by galaxy
mergers. There is no doubt that in the local universe there are
examples of galaxies merging with each other (e.g., \citealt{depropris2007}), 
which eventually will evolve into a system more massive than either of the
progenitors. During this process the physical evolution of the resulting
remnant galaxy will also change, as mergers are potentially the dominant 
method for triggering star formation and producing feedback by ejecting, or 
heating, gas (e.g., \citealt{cox2004}). Mergers can also spur on the 
formation of black holes and trigger AGN activity, as both observations
and theory have shown (e.g., \citealt{hopkins2008}; \citealt{bundy2008}).
While galaxy mergers are seen in the local and distant universe, the exact 
role of mergers in the formation and evolution of galaxies over cosmic 
history is still uncertain.

There are two primary methods for tracing the merger history of galaxies in 
observations: morphological identification techniques and the close galaxy 
pair method (e.g. \citealt{patton2000}). Examples of morphological techniques are the concentration--asymmetry--clumpiness method (CAS, hereafter, e.g. \citealt{cons2003a}) and the Gini-M20 method \citep{lotz2004}.
The CAS method identifies galaxies undergoing mergers 
on the basis of observed morphological properties (e.g., \citealt{cons2003a}) and is a non-parametric technique for measuring the shapes and structures of galaxies on resolved CCD images (e.g., \citealt{cons2000a}; \citealt{bershady2000}; \citealt{cons2002}; \citealt{cons2003a}).
All methodologies have been demonstrated to be internally consistent with each other (\citealt{depropris2007}; \citealt{cons2008}; \citealt{manfred2008}).
However, since each method identifies galaxies at a different stage during the merger, overall the galaxy populations investigated by each method can be substantially different.
The main uncertainty in all methods resides in the time-scale $\tau_{\rm m}$ used to identify galaxies undergoing mergers (e.g., \citealt{cons2006}; \citealt{lotz2008b}). Currently, the most likely time-scale to which the CAS and pair methods are sensitive is believed to be $\tau_{\rm m} \sim 0.4-1$ Gyr.

In the last decade a relatively large sample of galaxy mergers has been 
collated (\citealt{patton2000}; \citealt{patton2002}; \citealt{cons2003a}; 
\citealt{hernandez2005}; \citealt{bundy2005}; \citealt{bundy2006}; 
\citealt{cons2006}; \citealt{cons2006b}; \citealt{bundy2007}; 
\citealt{depropris2007}; \citealt{lotz2008a}; \citealt{cons2008}; 
\citealt{patton2008}; \citealt{bluck2009}) and the merger history of 
galaxies has been traced with sufficient accuracy to perform a 
detailed comparison with theoretical models.
Mergers are just now beginning to be identified at $z > 2$ through either 
close galaxy pairs (e.g., \citealt{bluck2009}), or structural methods 
(e.g., \citealt{cons2003b}; \citealt{cons2008}; \citealt{lotz2008b}).
What is becoming clear is that the merger fraction increases at larger 
look back times (higher redshifts). This increase can be parameterised as 
either a power-law, or a combined power-law/exponential 
(e.g., \citealt{carlberg1990}; \citealt{cons2006}; \citealt{cons2008}). 
These measured merger fractions in the early universe can be converted into 
merger rates up to $z \sim 3$, and based on these we can attempt to determine the role of major mergers in the formation of galaxies. 

In this work, we retrieve the predicted galaxy merger history evolution from
the Millennium-based semi-analytic model of \citet{bertone2007} as a 
function of stellar mass. We then compare the model predictions with a 
selection of observational results and examine separately the merger history 
of galaxies with stellar masses in the intervals $10^{9}$ \solm\ $<M_{\star}< 10^{10}$ \solm, $M_{\star} > 10^{10}$ \solm\ and $M_{\star} > 10^{11}$ \solm,  at $z < 3$.
Given the uncertainty in the observational results introduced by the 
time-scale $\tau_{\rm m}$, in this paper we assume $\tau_{\rm m} = 0.4$ Gyr 
as our fiducial value and we investigate how results can change for a longer
time-scale of $\tau_{\rm m} = 1$ Gyr.

In general, we find that the predicted merger history varies as a function
of stellar mass, redshift and merger mass ratio. The predicted major and minor merger rates, that is the number of mergers per unit volume and unit time, increase with lower stellar masses. We also find that the predicted merger rate increases with redshift and reaches a plateau at $z<1$ for galaxies with $M_{\star} > 10^{11}$ \solm. The predicted merger rate for less massive galaxies decreases with redshift. At $z < 1.5$ we find a good agreement between predicted and observed merger rates. However, the predictions for the merger fractions match the observations only for the most massive galaxies with $M_{\star} > 10^{11}$ \solm\ at $z < 2$. The predicted major merger fractions are about 10 times smaller than the observational estimates for galaxies with $M_{\star} > 10^{10}$ \solm\ at $z \sim 0.5$. We discuss the various ways in which the merger fraction predictions could be incorrect and conclude that the simulations might not predict enough mergers between galaxies with $M_{\star} < 10^{11}$ \solm. This could potentially account for other problems with matching observations to semi-analytical models, including the fact that there are more massive galaxies at high redshift than predicted and the satellite problem (e.g., \citealt{moore1999}; \citealt{cons2007}).

This paper is organised as follows. In \S \ref{data} we describe the 
observational data used to compare with the theoretical predictions. In 
particular, we briefly discuss our criteria for selecting the observational sample and some 
basic properties of the galaxies in the sample, such as their morphology 
and stellar masses. \S \ref{millennium} describes the Millennium galaxy 
catalogue of \citet{bertone2007} and how the merger rates and fractions have 
been extracted from the simulation. In \S \ref{compare} we present the 
model results and compare them to the observational data. Finally, \S 
\ref{discussion} discusses our findings and \S \ref{conclusion} summarises 
our results.

\section{Merger History Data}
\label{data}

In this Section we give a brief description of the data sets used for comparison with the Millennium simulation (Subsections \ref{sec21} and \ref{sec22}) and of how the merger fractions and rates are measured (Subsection \ref{definition}).

We use results from a series of works, based on both the structural asymmetries of galaxies and pair counts, that investigate the merger history for galaxies as a function of stellar mass (\citealt{cons2003a}; \citealt{depropris2007}; \citealt{cons2008}; \citealt{bluck2009}; \citealt{cons2009}).
Interested readers should examine these papers for the many details involved.
The observational quantities we present in this paper are measured using a 
standard $\Lambda$CDM cosmology with $H_{0} = 70$ \kms\ Mpc\1\ and $\Omega_{\rm m} = 1 - \Omega_{\lambda}$ = 0.3. This differs slightly from the cosmology
used within the simulation, although the effects of this are minor.

The first step to investigate the merger history of galaxies requires selecting a robust sample of mergers. The observed merger history must be constructed using different data sets, as the ability to measure merger properties at different redshifts requires different observing conditions and telescopes. This is due, e.g., to the fact that to obtain a large sample of nearby galaxies to measure the local merger rates and fractions requires a wide area survey, while detecting galaxies at high redshift requires very deep exposures, typically with the Hubble Space Telescope (e.g. \citealt{cons2008}; \citealt{lotz2008a}).

\subsection{CAS-Selected Mergers}
\label{sec21}

At low redshifts, we use datasets that estimate the merger quantities through the CAS method. The basic idea behind structural methods is that galaxies have light distributions that reveal their past and present formation modes \citep{cons2003a}.
The classical rest-frame optical CAS definition for determining whether a system is undergoing a merger is given by the two conditions \citep{cons2003a}:

\begin{equation}\label{cond_one}
A > 0.35\,{\rm and}\, A > S,
\end{equation}

\noindent that is, the asymmetry $A$ must be larger than a threshold value, and must exceed the value of the clumpiness $S$ of the galaxy. These selection criteria have been tested against samples of nearby galaxies (\citealt{cons2003a}; \citealt{depropris2007}) and N-body simulations \citep{cons2006}. Tests at higher redshifts are limited to either small samples \citep{cons2008}, or to the GEMS survey \citep{jogee2009}.
The CAS method empirically finds systems which are nearly all major mergers (e.g., \citealt{cons2000b}; \citealt{cons2003a}; \citealt{hernandez2005}; \citealt{cons2006}), although there are hints that in some cases the mergers identified by CAS could be contaminated by mergers with a smaller mass ratio \citep{jogee2009}.

For galaxies at $z \sim 0.05$, which is effectively our nearby galaxy merger fraction, we use the results from \citet{depropris2007}. \citet{depropris2007} estimate the merger fraction using both the number of systems in pairs and their morphological structure as identified by CAS.
At $0.2 < z< 1.2$ we take the merger fractions measured for a combined COSMOS and Extended Groth Strip sample by \citet{cons2009}, who consider more than $20,000$ galaxies with $M_{\star} > 10^{10}$ \solm.
The merger quantities of systems at $1 < z < 2$ are from a combined sample of Hubble Deep Field (HDF) and Hubble Ultra-Deep Field (UDF) galaxies \citep{cons2008}.
The compilation of observational results we use in this work is in good agreement with other published results (e.g. \citealt{patton2002}; \citealt{lin2004}; \citealt{kartaltepe2007}; \citealt{lin2008}; \citealt{patton2008}). A detailed comparison between the different datasets is presented in \citet{cons2009} and we refer the reader to that paper for more information.

\subsection{Mergers from Pair Methods}
\label{sec22}

At the highest redshifts we consider, we use estimates of the merger quantities based on pair counts. The pair method is a complementary and independent 
approach for finding the merger history of galaxies (see \citealt{cons2008} for details).

In the pair count method, close pairs are defined as galaxy systems with a physical separation of less than 30 h$^{-1}_{70}$ kpc. Major merger pairs are further defined as systems with galaxies with magnitudes within $\pm 1.5$ of each other.
The pair fraction, defined as the number of galaxy pairs divided by the number of galaxies within the sample, is not a direct measure of the merger fraction. The conversion of a pair fraction in to a merger fraction (e.g. \citealt{depropris2007}; \citealt{cons2008}) is affected by large uncertainties and in particular suffers from uncertainties in the determination of the time-scale for merging. We will discuss the time-scale to which the CAS structural method and the pair method are sensitive in the next Sections. This time-scale is also required to convert fractions to merger rates and to effectively compare the observed results with models.
In this work we derive merger fractions and rates from pairs only at $z>2$ based on \citet{bluck2009} drawn from galaxy pairs identified in deep NIR imaging from the GOODS NICMOS Survey.

\subsection{Merger fractions and merger rates}
\label{definition}

In this Subsection we define the main quantities we will be investigating in the following, namely the merger fraction and the merger rate, and we discuss possible biases in their estimation both in observations and simulations.

The merger fraction is the number of mergers $N_{\rm m}$ within a given redshift bin and stellar mass limit, divided by the number of galaxies $N_{\rm T}$ within the same redshift $z$ and stellar mass bin:
\begin{equation}\label{mergfrac}
f_{\rm m}\left( {\rm M_{\star}},z\right) = \frac{N_{\rm m}\left( {\rm M_{\star}},z\right)}{N_{\rm T}\left( {\rm M_{\star}},z\right)}.
\end{equation}
The conversion of the merger fraction, that is the ratio of the number of mergers to the total number of galaxies, into a galaxy merger fraction $f_{\rm gm}$ \citep{cons2006}, which is the number of galaxies merging divided by the
total number of galaxies, can be done through:   
\begin{equation}\label{galmergfrac}
f_{\rm gm} = \frac{2 \times f_{\rm m}}{1+f_{\rm m}}.
\end{equation}
If each pair produces a merger, the pair fraction is equivalent to the merger fraction, sans different time-scales, and the value of the galaxy merger fraction $f_{\rm gm}$ is roughly twice that of the merger fraction $f_{\rm m}$, defined in Eq. (\ref{mergfrac}) \citep{cons2006}. 

The conversion of pair and merger fractions into merger rates requires knowledge of the time-scale $\tau_{\rm m}$ to which the CAS and the pair methods are sensitive.
This time-scale is critically important, because it ultimately determines the values of the merger rates and fractions both in simulations and observations.
Simulations in general, including the Millennium, have well-defined merger rates, that can be measured accurately. However, the merger fractions depend on the time-scale over which the number of mergers are counted.
In fact, the fraction of galaxies that have merged within a given time interval is roughly proportional to the time-scale considered. However, the number of mergers per unit time is independent of the time-scale used.

The merger rate, defined as the number of galaxies merging per unit time and per unit volume, which we denote as \rate, is given by:
\begin{equation}\label{mergerrate}
\Re(z) = \frac{f_{\rm gm}(z) \cdot n_{\rm gm}(z)}{\tau_{\rm m}},
\end{equation}
where $n_{\rm gm}$ is the number density of galaxies within a given stellar 
mass range.
Although the observational value of $\tau_{\rm m}$ is not well constrained and there are a few indications that it might not be the same for structural and pair methods (\citealt{lotz2008a}; \citealt{manfred2008}), in general it varies between about 0.4 Gyr and 1 Gyr (e.g., \citealt{cons2006}; \citealt{lotz2008b}).
In this work we assume two different values for $\tau_{\rm m}$, that is $\tau_{\rm m}=0.4$ Gyr and $\tau_{\rm m}=1$ Gyr, in order to take in to account at first order the possible systematics involved with comparing data with models.
The value of $\tau_{\rm m}=0.4$ Gyr is based on the sensitivity of the CAS method, which is believed to identify galaxies that have undergone mergers within approximately such a time-scale. The time-scale of 1 Gyr has been suggested by \citet{lotz2008b} on the basis of results of hydro-dynamical simulations.

Another issue to address while comparing models with simulations is the minor vs. major merger sensitivity of different methods. Pair fractions require galaxies to have magnitudes within $\pm 1.5$ of each other, but do not provide exact information about the respective galaxy masses. Although it is usually assumed that the mass ratio in galaxy pairs is larger than about 1:4, there might be large variations from pair to pair. As discussed in Section \ref{sec21}, the CAS method identifies galaxies that have already merged. This implies that we do not know the progenitor masses, although simulations predict that minor mergers with mass ratios 1:10 or less do not produce significant structural asymmetries (\citealt{hernandez2005}; \citealt{cons2006}).
The limited information available suggests that the CAS method only picks out major mergers with a mass ratio of about 1:4 or greater, but does not exclude contamination from mergers with smaller mass ratio \citep{jogee2009}.

\section{The Millennium galaxy catalogue}
\label{millennium}

The numerical results presented in this work use the galaxy catalogues of \citet{bertone2007}, publicly available from the Millennium 
website\footnote{http://www.mpa-garching.mpg.de/galform/virgo/millennium}.
The Millennium simulation \citep{springel2005} uses a $\Lambda$CDM cosmology 
with parameters $\Omega_{\rm m}=0.25$, $\Omega_{\rm b}=0.045$, $h=0.73$, 
$\Omega_\Lambda=0.75$, $n=1$ and $\sigma_8=0.9$. The Hubble constant is 
parameterised as $H_{\rm 0} = 100$ $h^{-1}$ km s$^{-1}$ Mpc$^{-1}$. The Millennium cosmology is slightly different from that assumed by the observations, but the uncertainties that may derive from this discrepancy should be negligible in comparison with the uncertainties introduced, for example, by the merging time-scale.  For example, at the highest redshifts the ratio of volumes for
the model cosmology is 1.1\% higher than the cosmology used in the
observations. This volume effect only influences the calculation of the
merger rates, and does not affect the calculation of the merger
fractions.

The simulation follows the evolution of about 20 million galaxies in a region of 500 \hm Mpc on a side. The galaxy catalogues were created using a variation of the semi-analytic model of \citet{delucia2007}, in which supernova feedback is implemented using a dynamical treatment of galactic winds \citep{bertone2005}. The model improves the treatment of galaxies with stellar masses 
$M_{\star} < 10^{11}$ \solm, but somewhat overpredicts the abundance of 
the most massive galaxies with $M_{\star} > 10^{11}$ \solm. For a 
detailed description of the model, we refer the interested reader to the original paper of \citet{bertone2007}. 

The merger history of the simulated galaxies is recovered by flagging galaxies that have undergone mergers and by saving additional information about the time when the mergers occur. In order to estimate the merger fractions and the merger rates as closely as possible to the observations, we count the number of galaxies $N_{\rm m}$ that have merged at least once within a time-scale $\tau_{\rm m}$ at redshifts $z=0$, 0.5, 1, 1.5, 2 and 3. The merger fractions are then calculated as in Eq. (\ref{mergfrac}) and the merger rates as in Eq. (\ref{mergerrate}).

The numerical resolution of the Millennium simulation allows us to correctly 
track the merging history of galaxies only for $M_{\star} > 10^{9.5}$ M$_{\sun}$. Accordingly, our results are not fully reliable below this mass threshold. Nonetheless, in a few cases we have chosen to 
display results for the merger rates and fractions in the interval 
$10^9$ M$_{\sun}<M_{\star} < 10^{10}$ M$_{\sun}$: the inability of the 
model to track the merger history of galaxies with $10^9$ M$_{\sun}<M_{\star} < 10^{9.5}$ M$_{\sun}$ translates in to lower predicted values than in the ideal case where the merger history is fully accounted for. This effect is certainly stronger for minor merger counts, but may also affect major mergers.

% Median Merging Times
\begin{figure}
\includegraphics[width=8.4cm]{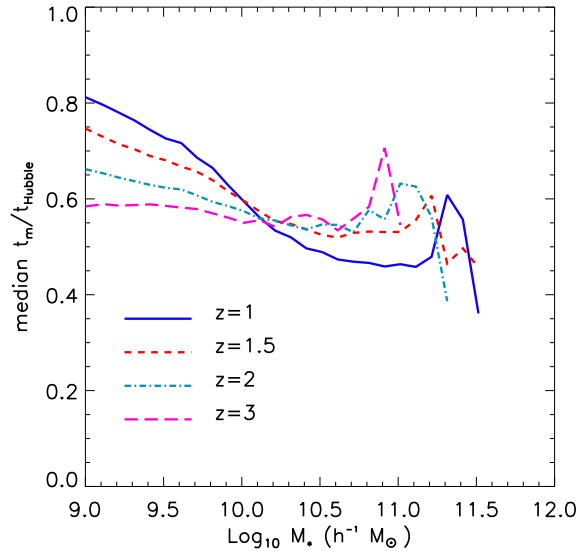}
\caption{The median merging time $t_{\rm m}$ in units of the Hubble time $t_{\rm Hubble}$ in the Millennium simulation as a function of the stellar mass of the satellite galaxy. Results are shown for $z=1, 1.5, 2$ and 3.}
\label{mtime}
\end{figure}

In the Millennium semi-analytic model, galaxies are classified as central or satellite galaxies \citep{springel2001}. Central galaxies are usually the most massive galaxies in haloes, while satellite galaxies are smaller objects  accreted through merging of dark matter haloes. When a smaller halo falls into a larger one, its galaxies become satellite galaxies of the resulting halo. Satellite galaxies can in turn be classified as satellites of type 1, that is galaxies associated with a dark matter substructure, or of type 2, that is galaxies whose dark matter substructure has completely merged with the central halo and can no longer be identified.

In the model, the merging time of galaxies, $t_{\rm m}$, is the lifetime of a satellite of type 2 and is defined as the time-scale between when a galaxy loses its dark matter substructure and the moment when it merges with the halo central galaxy \citep{springel2001}. The merging time is given by the dynamical friction time-scale \citep{binney1987} and depends on both the virial properties of the dark matter substructure to which the galaxy was previously attached and on those of the central halo.

This definition of merging time is different from the ``total merging time'', which is instead the time between the moment a galaxy falls into a larger halo, and when it actually merges with the halo central galaxy, independently of when its associated substructure merges with the cluster halo. The total merging time is the sum of the merging time $t_{\rm m}$ plus the time that a galaxy spends as a satellite of type 1.
Typically, the time spent by a galaxy as a satellite of type 1 is much shorter than the time spent as a type 2. The time-scale for a substructure to dissolve in to a larger halo varies, e.g., with the parent halo and with the substructure dark matter mass, but on average it is shorter than 1 Gyr. As such, the time spent by a satellite as a type 1 is about an order of magnitude shorter than the merging time $t_{\rm m}$.

Fig. \ref{mtime} shows the median merging time of galaxies $t_{\rm m}$ as a function of stellar mass at $z=1, 1.5, 2$ and 3.
The median merging time t$_{\rm m}$ depends only weakly on stellar mass at $z=3$, but at later times the more massive galaxies tend to have shorter merging times, on average by a factor of two, compared with lower mass galaxies.
There is, in addition, a strong evolution with redshift. In particular, 
$t_{\rm m}$ increases steadily with decreasing $z$, and it is about 3 times 
larger at $z=1$ than at $z=3$. The spikes at the highest stellar masses are 
the result of poor statistics, since only a handful of galaxies populate 
these mass bins.

The merging time-scale in the semi-analytic model depends on the virial masses 
of the merging galaxies, but not on their baryon content. Therefore, because the models of \citet{bertone2007} and \citet{delucia2007} use the same merger trees, galaxies always merge at the same time in both models. However, since 
these two models do not predict the same stellar mass, our results for the merger 
fraction and merger rates as a function of stellar mass and redshift might differ from those of \citet{delucia2007}. We discuss the difference between the two model predictions in Subsection \ref{varymodel}.

\begin{figure}
\includegraphics[width=8.4cm]{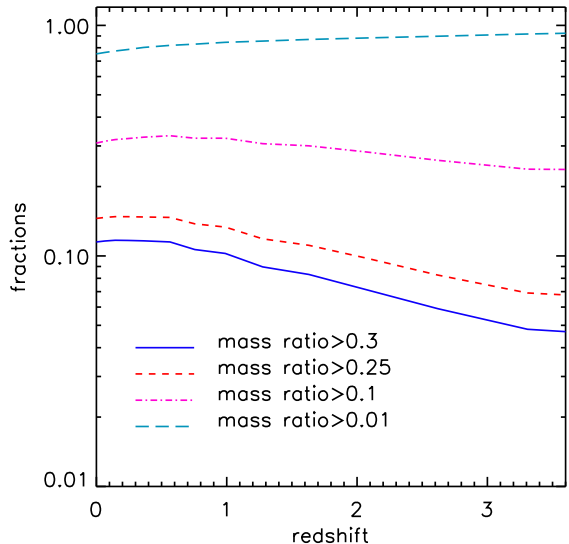}
\caption{The fraction of galaxy mergers with a mass ratio larger than 0.3 (solid line), 0.25 (dashed line), 0.1 (dot-dashed line) and 0.01 (long dashed line) as a function of redshift, for simulated galaxies with $M_{\star} > 10^9$ \solm.}
\label{mass}
\end{figure}

In the semi-analytic model the mass ratio over which a merger is defined as major is assumed to be $m_{\rm satellite} / m_{\rm central} > 0.3$, where $m_{\rm satellite}$ is the mass in baryons of the merging galaxy and $m_{\rm central}$ is the mass of the central galaxy.
This may not be fully consistent with the observations, where best estimates 
suggest that major mergers are those with mass ratios larger than $\sim 0.25$. This is quantified in Fig. \ref{mass}, which shows the predicted fraction of mergers with mass ratios larger than 0.3, 0.25, 0.1 and 0.01 as a function of redshift, for galaxies with $M_{\star} > 10^9$ \solm. While almost all mergers have a mass ratio larger than 0.01, the fraction of mergers with larger mass ratios decreases quickly. In particular, we find that mergers with a mass ratio larger than 0.25 are about 20 per cent more numerous than mergers with a mass ratio of 0.3. This implies that our estimates of the merger fraction and merger rate may be 20 per cent lower than the observations simply because of the Millennium simulation assumption of a higher mass threshold for defining major mergers. We will discuss this further in Sec. \ref{discussion}.

\section{Results}
\label{compare}

In this Section we present our results for the merger history of galaxies in the Millennium simulation. We first present results of predicted merger fractions and rates as a function of stellar mass in Subsection \ref{basic} and we then proceed to compare the simulation predictions with the observational results in Subsections \ref{fracs}, \ref{param} and \ref{rate_evolution}.

\subsection{Basic Features of the Predicted Merger History}
\label{basic}

\begin{figure*}
\includegraphics[width=0.48\textwidth]{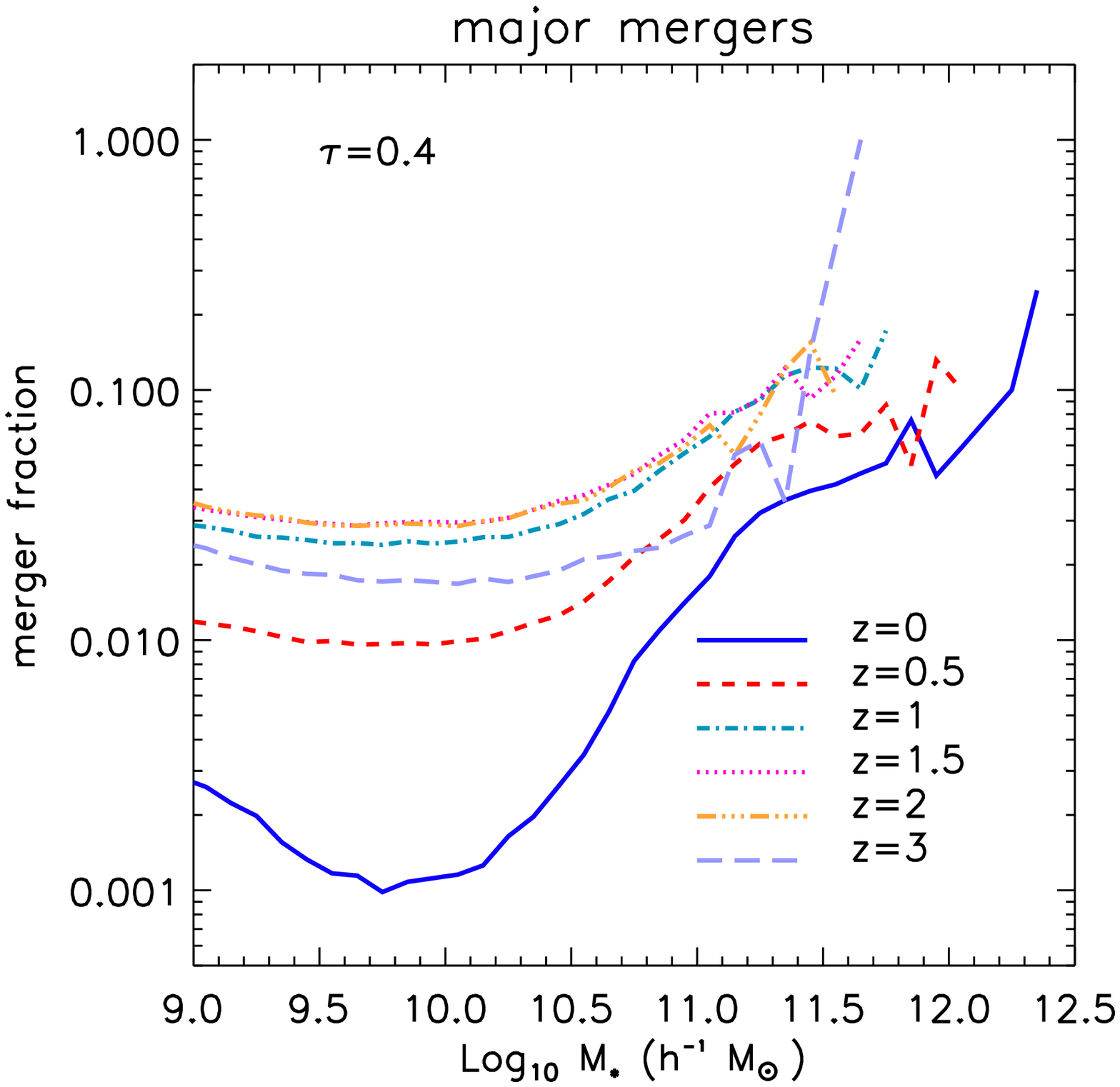}
\includegraphics[width=0.48\textwidth]{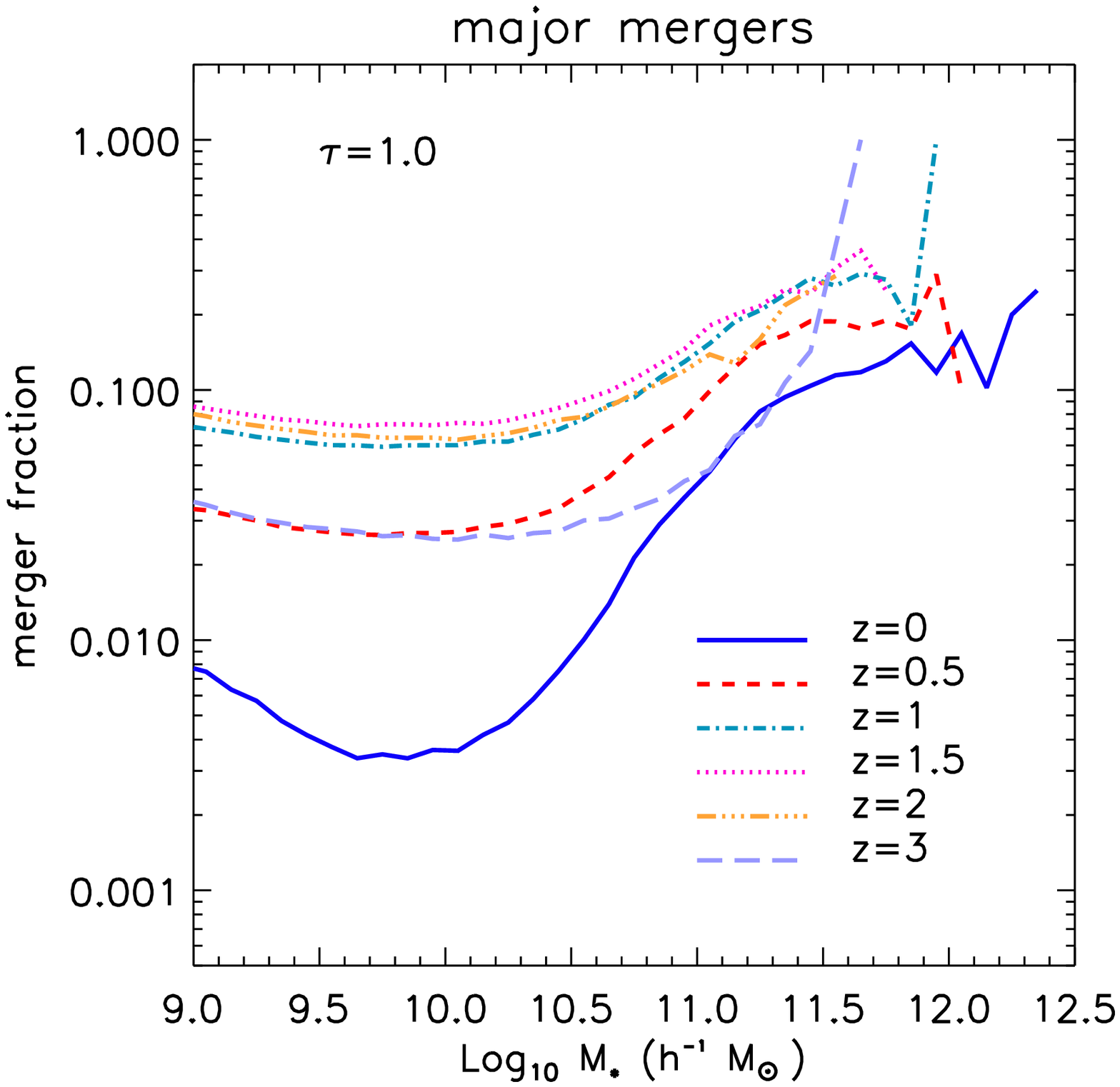}
\includegraphics[width=0.48\textwidth]{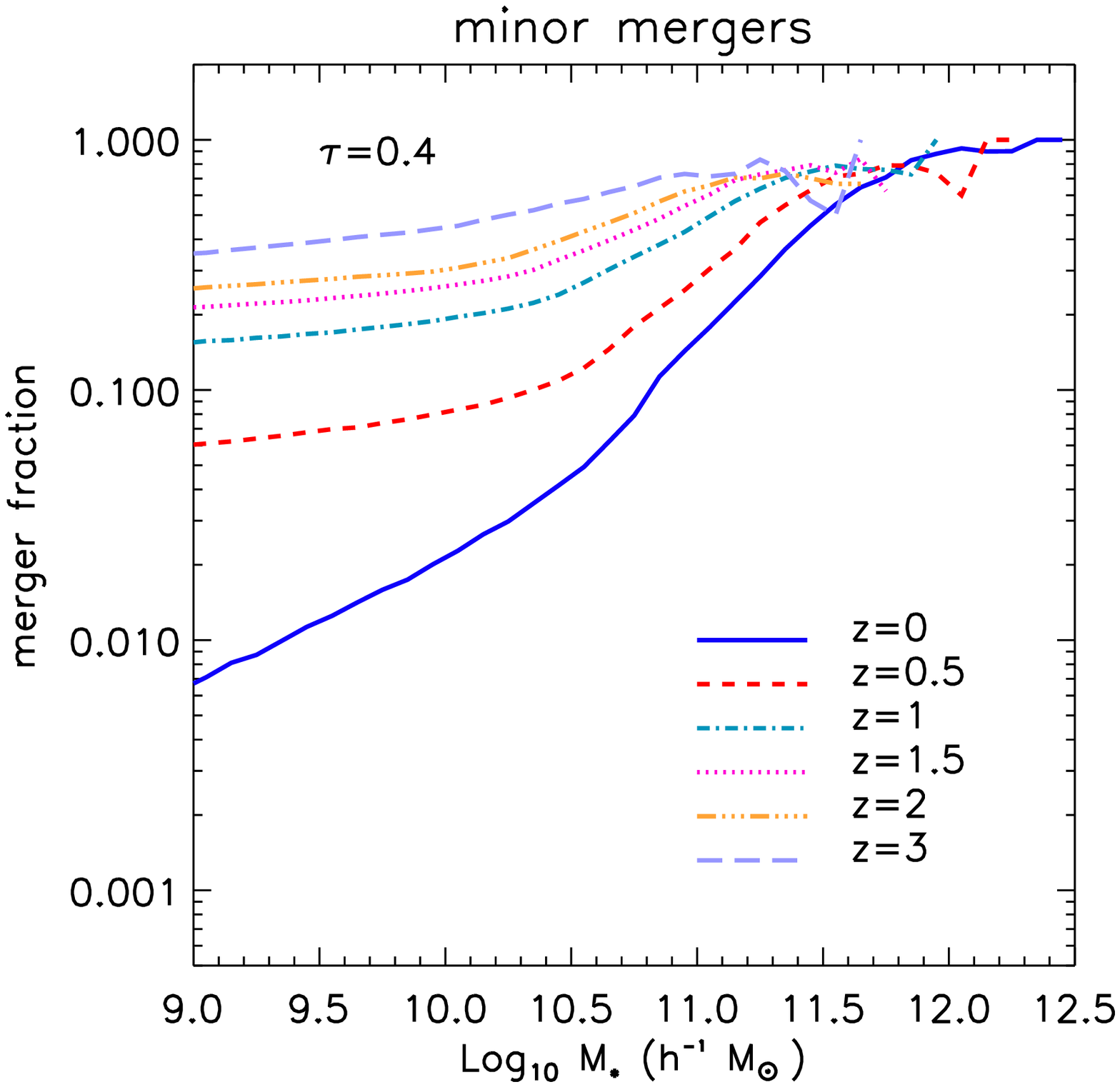}
\includegraphics[width=0.48\textwidth]{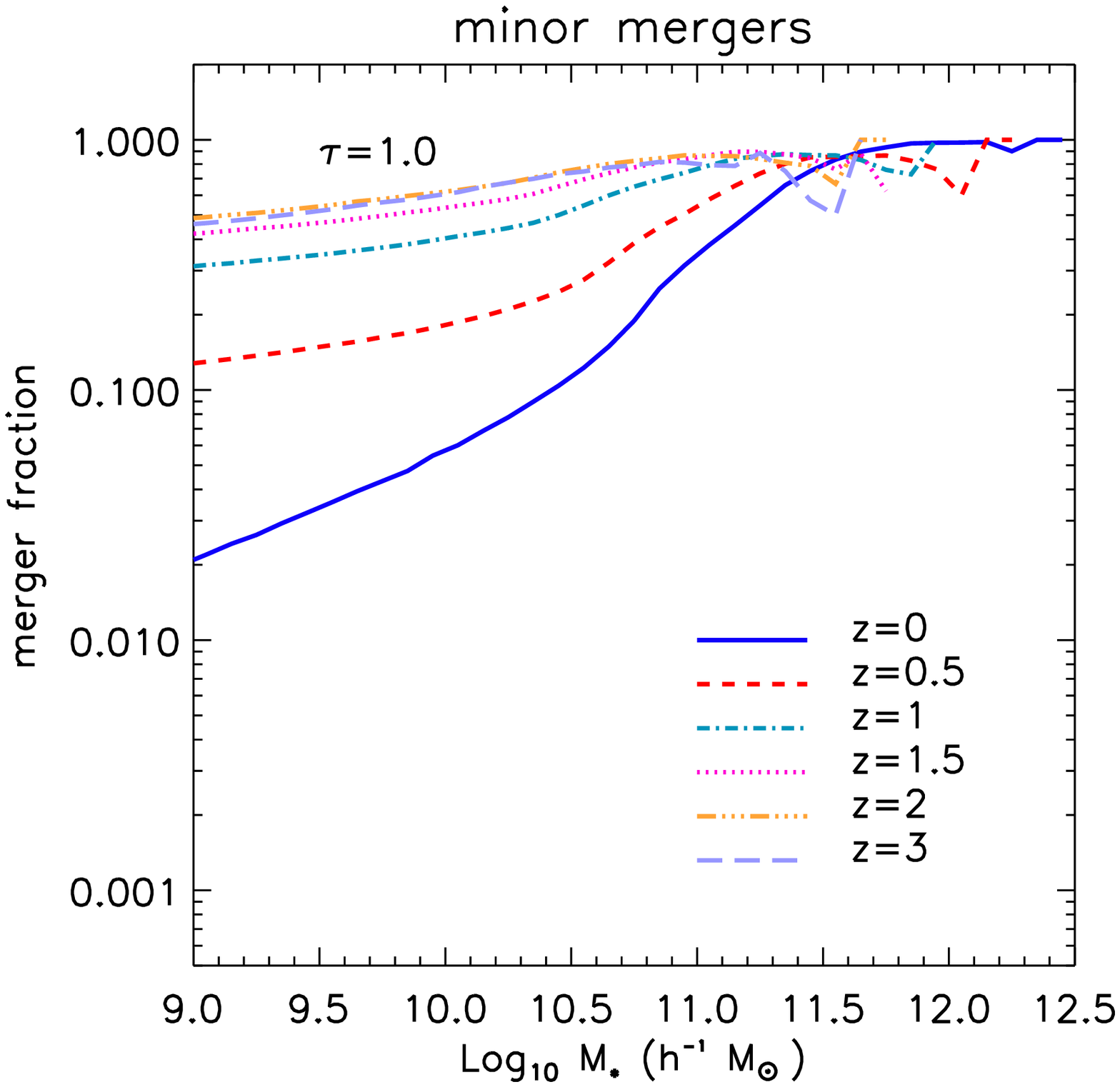}
\caption{Predictions for the merger fractions of galaxies as a function of stellar mass and redshift in the Millennium simulation. The left panels show the merger rates calculated for a time-scale $\tau_{\rm m} = 0.4$ Gyr, while the right panels assume a time-scale $\tau_{\rm m} = 1.0$ Gyr. The merger fractions are shown separately for minor mergers (lower panels) and major mergers (upper panels).}
\label{fig2}
\end{figure*}

\begin{figure*}
\includegraphics[width=0.48\textwidth]{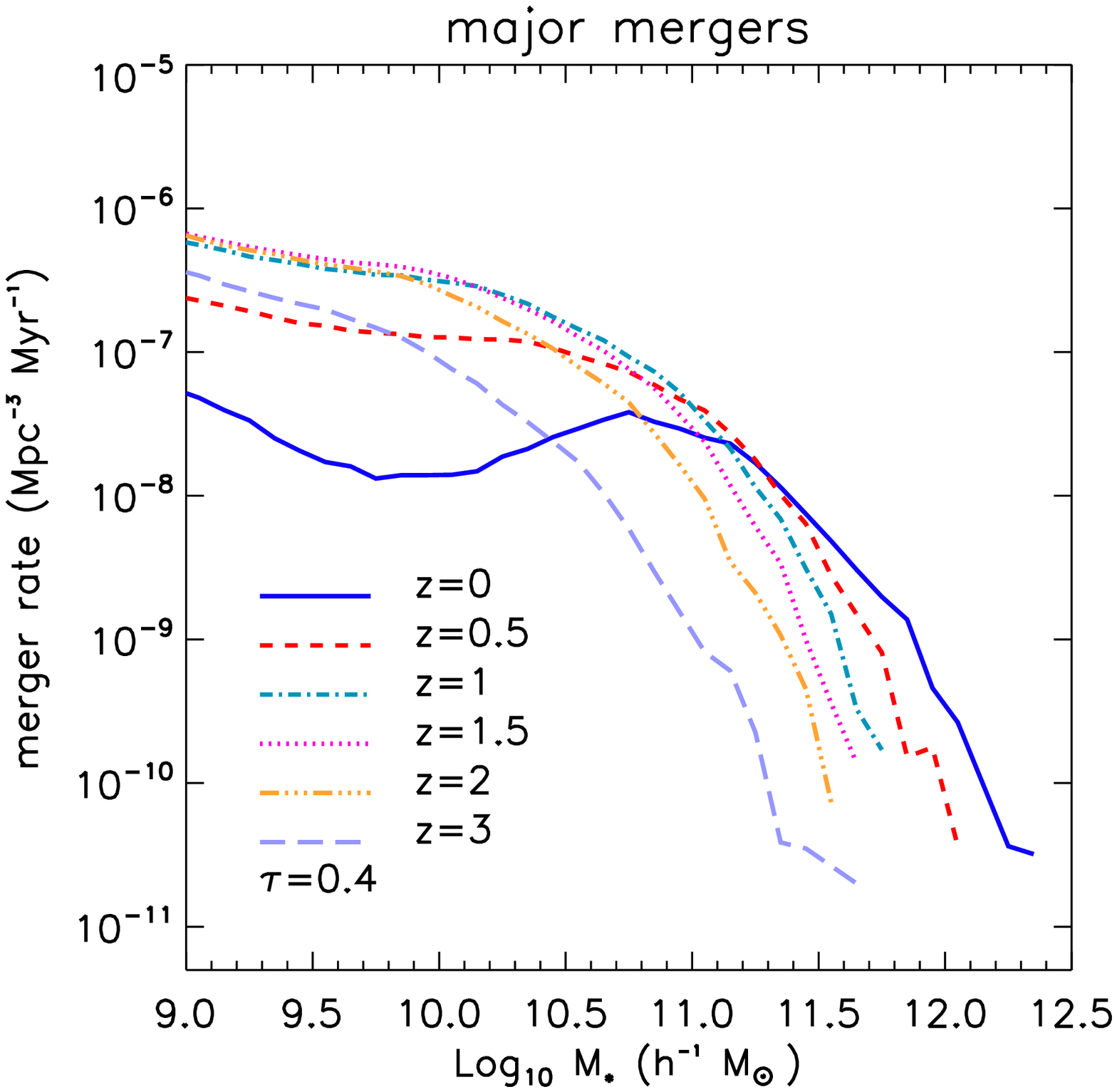}
\includegraphics[width=0.48\textwidth]{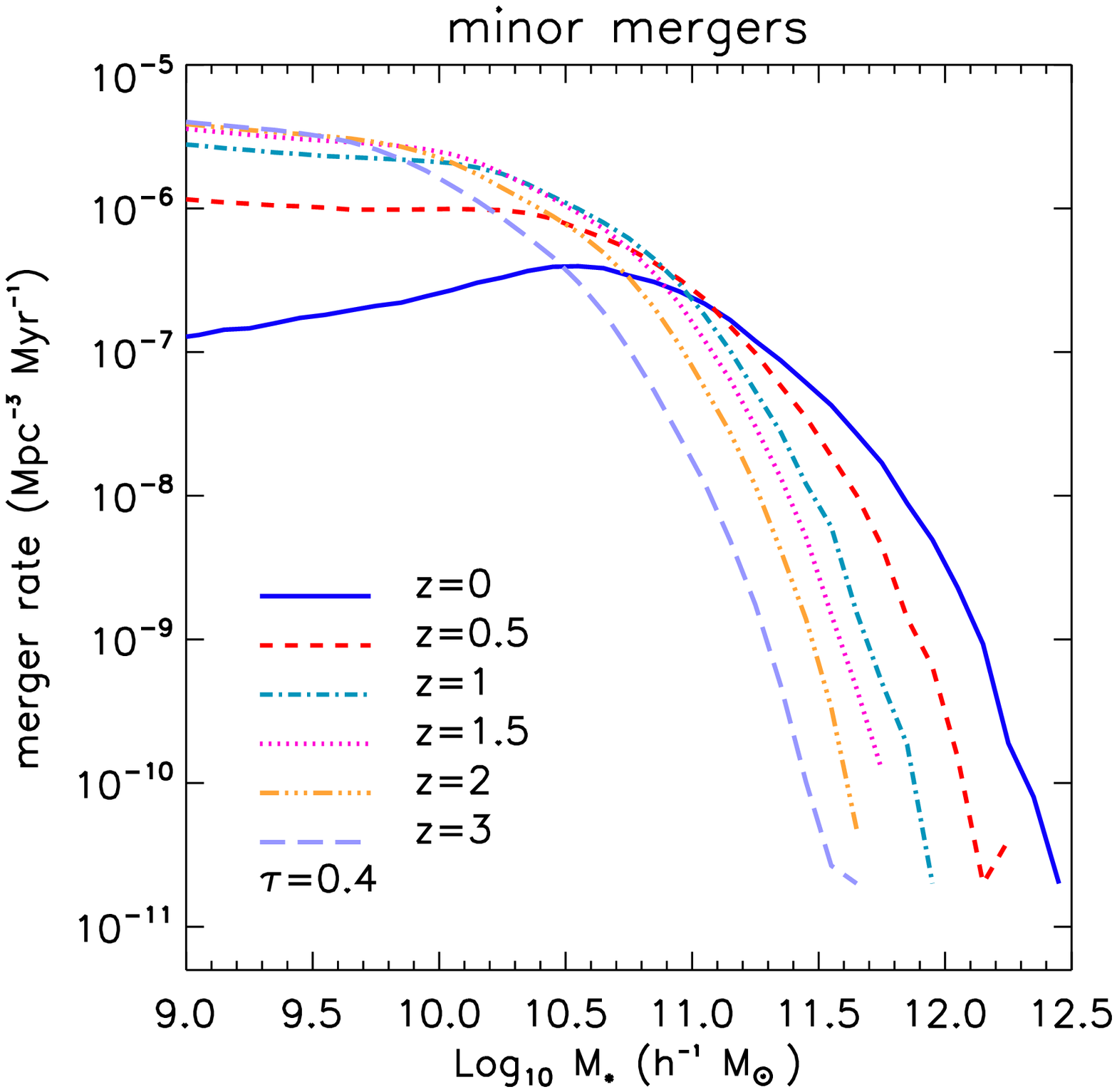}
\caption{The merger rate in the Millennium simulation as a function of stellar mass. The left panel shows the major merger rate and the right panel the minor merger rate. In both panels, each line shows results at different redshifts.}
\label{fig0}
\end{figure*}

In this Subsection we present results on the evolution of the merger fraction and the merger rate as a function of stellar mass and redshift in the Millennium simulation.

Fig. \ref{fig2} shows the predicted evolution of the merger fraction with redshift as a 
function of stellar mass for both major (upper panels) and minor mergers 
(lower panels). 
To investigate the dependence of the merger fraction on the time-scale 
over which mergers are counted, we show results for two cases: one in which 
we have considered mergers that have occurred within a time-scale $\tau_{\rm m}=0.4$ Gyr (left panels) and a second case with a time-scale $\tau_{\rm m}=1$ Gyr (right panels).
We find little difference between the results obtained using the two different time-scales. Both the shape and the general history of the minor and major mergers are very similar at high-z, making the number of mergers per unit time evolving slowly.

There is a slight difference between the predicted evolution of the merger fraction at high and low stellar masses, with a change in slope at about $M_{\star} \sim 10^{10.5}$ \solm. The major merger fraction is lowest 
at low redshift (blue solid line) and is highest for the most massive 
systems with $M_{\star} > 10^{11}$ \solm\ at all redshifts. At high redshifts there are few galaxies with stellar masses larger than $M_{\star} \sim 10^{11}$ \solm\ and this result is affected by small number statistics. Within the observed sample, there are no galaxies with $M_{\star} > 10^{12}$ \solm\ and very few are found at $M_{\star} > 10^{11.5}$ \solm.
The predicted merger fraction for galaxies with $M_{\star} \sim 10^{10}$ \solm\ at $z = 0$ is $f_{\rm m} < 0.01$ for mergers within a time-scale $\tau_{\rm m} = 0.4$ Gyr. This result is roughly consistent with what is found when examining the nearby merger fraction history (e.g., \citealt{depropris2007}. 

The minor merger fractions shown in the lower panels of Fig. \ref{fig2} display a similar trend to the major mergers, plus a few additional interesting features. The first is that minor mergers are very common in massive galaxies, with nearly all galaxies at $M_{\star} > 10^{11}$ \solm\ having a minor merger occurring within the past 0.4 Gyr at every redshift.
At lower stellar masses, there is an increasingly large difference in the minor merger fraction evolution, with a high ratio of $f_{\rm m,minor} > 0.3$ at $z = 3$, which declines rapidly with redshift by an order of magnitude at $z=0$. This means that in the Millennium simulation less massive galaxies undergo many more minor mergers at high redshift than in the recent past. 

One could argue that if the mass resolution of numerical simulations could be increased indefinitely, the minor merger fraction would be unity for all galaxies. This is certainly true if one considers as minor mergers all accretion events, independently of the mass ratio between the central galaxy and the infalling cloud.
As we show in Fig. \ref{mass}, between 80 and 90 per cent of mergers in the Millennium simulation have mass ratios larger than 0.01. This mass ratio is a small number and includes accretion of small satellites that do not affect the morphology of the central galaxy and would not be found as members of galaxy pairs.
A fraction of these merging satellites have stellar masses below the resolution of the simulation, especially when galaxies with $M_{\star} < 10^{10}$ \solm\ are considered. However, above $M_{\star} > 10^{10}$ \solm, most events we consider as minor mergers have a non-negligible mass ratio. Imposing a minimum value for the mass ratio for mergers to be considered in our analysis, as for example a minimum mass ratio of 0.05, would modify the minor merger results by about a factor of 2 for the most massive galaxies, but would not change the qualitative behaviour of the minor merger fractions and rates.
The redshift evolution of the minor merger fraction as a function of stellar mass shown in Fig. \ref{fig2} is an indication that resolution effects alone do not shape the distribution of the minor merger fractions as a function of stellar mass. According to Fig. \ref{mass}, major mergers represent between 5 and 10 per cent of all merger events in the Millennium.

% merger rates vs. stellar mass

Fig. \ref{fig0} shows predictions of the merger rate as a function of stellar mass in units of Mpc$^{-3}$ Myr$^{-1}$. The evolution of the merger rate with redshift is shown for both major mergers (left panel) and minor mergers (right panel). The different lines indicate results at different redshifts.
The merger rate tends to be higher for less massive galaxies and lower for the most massive galaxies.
The shape of the merger rate as a function of stellar mass mostly reflects the shape of the stellar mass function $n_{\rm gm} (z)$, which enters the definition of the merger rate according to Eq. \ref{mergerrate}. In fact, while the number density of galaxies varies by up to five orders of magnitude for $10^9$ \solm $<M_{\star} < 10^{12}$ \solm systems, the merger fractions vary by at most a factor of 100, as shown in Fig. \ref{fig2}.
The major and the minor merger rates are fairly constant for galaxies with $M_{\star} < 10^{11}$ \solm, at $z > 1$ and decrease with time for $z < 1$.
The merger rates of galaxies with $M_{\star} > 10^{11}$ \solm\ strongly decrease with redshift. This partially reflects the increase in the number density of galaxies with time.

\subsection{Comparing the Observed and Predicted Merger Fraction Evolution}
\label{fracs}

\begin{figure*}
\includegraphics[width=0.43\textwidth]{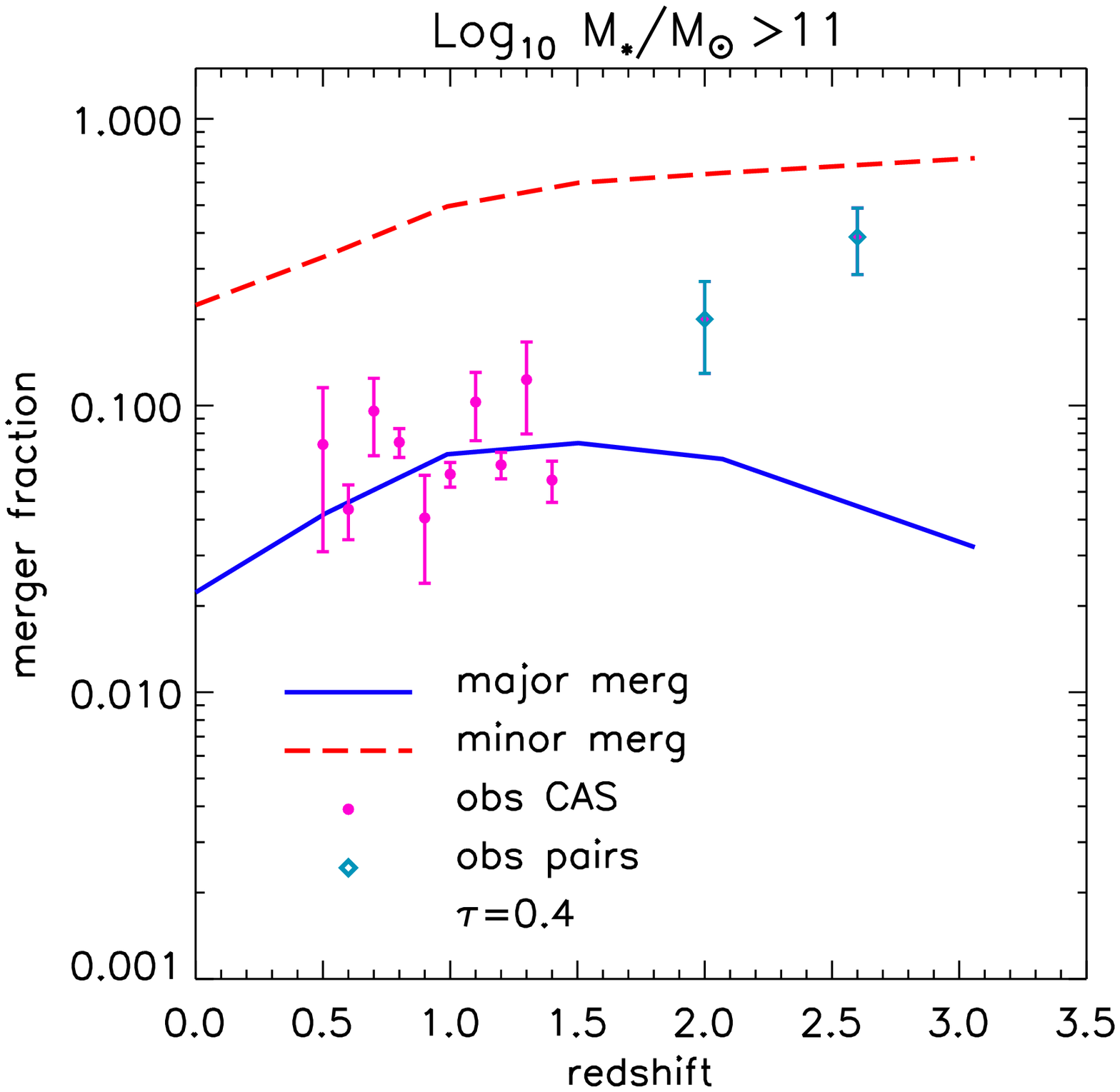}
\includegraphics[width=0.43\textwidth]{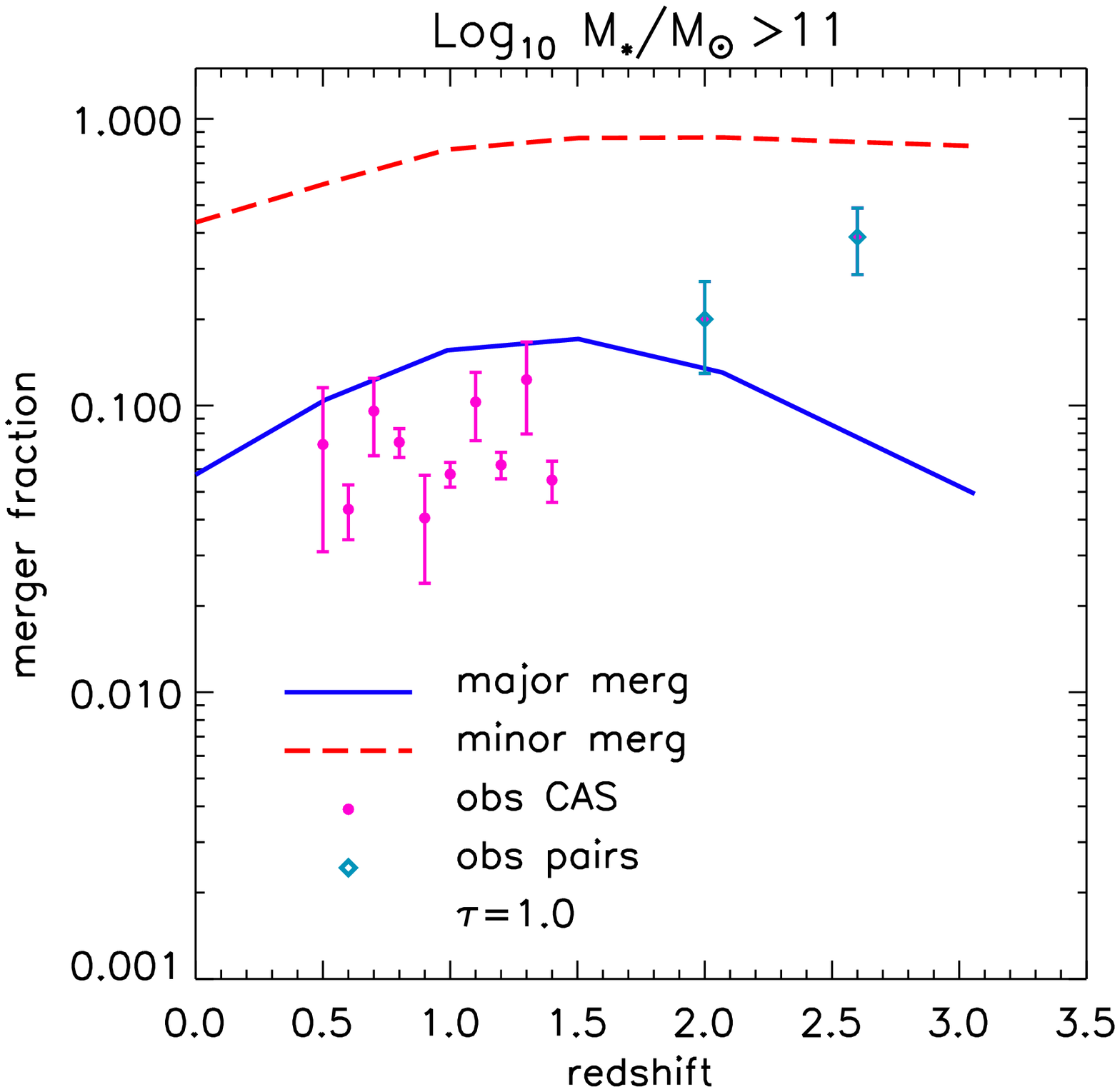}
\includegraphics[width=0.43\textwidth]{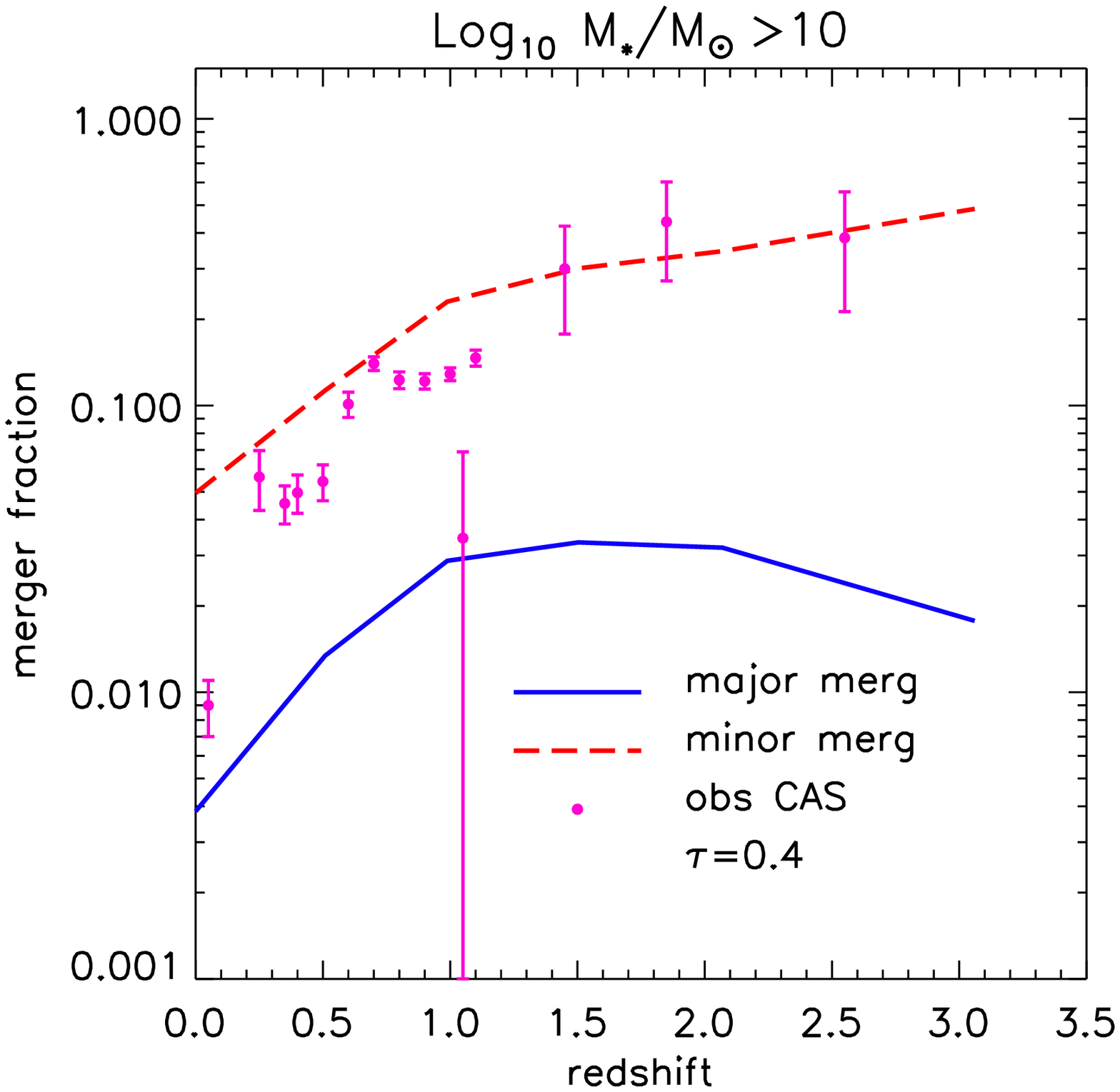}
\includegraphics[width=0.43\textwidth]{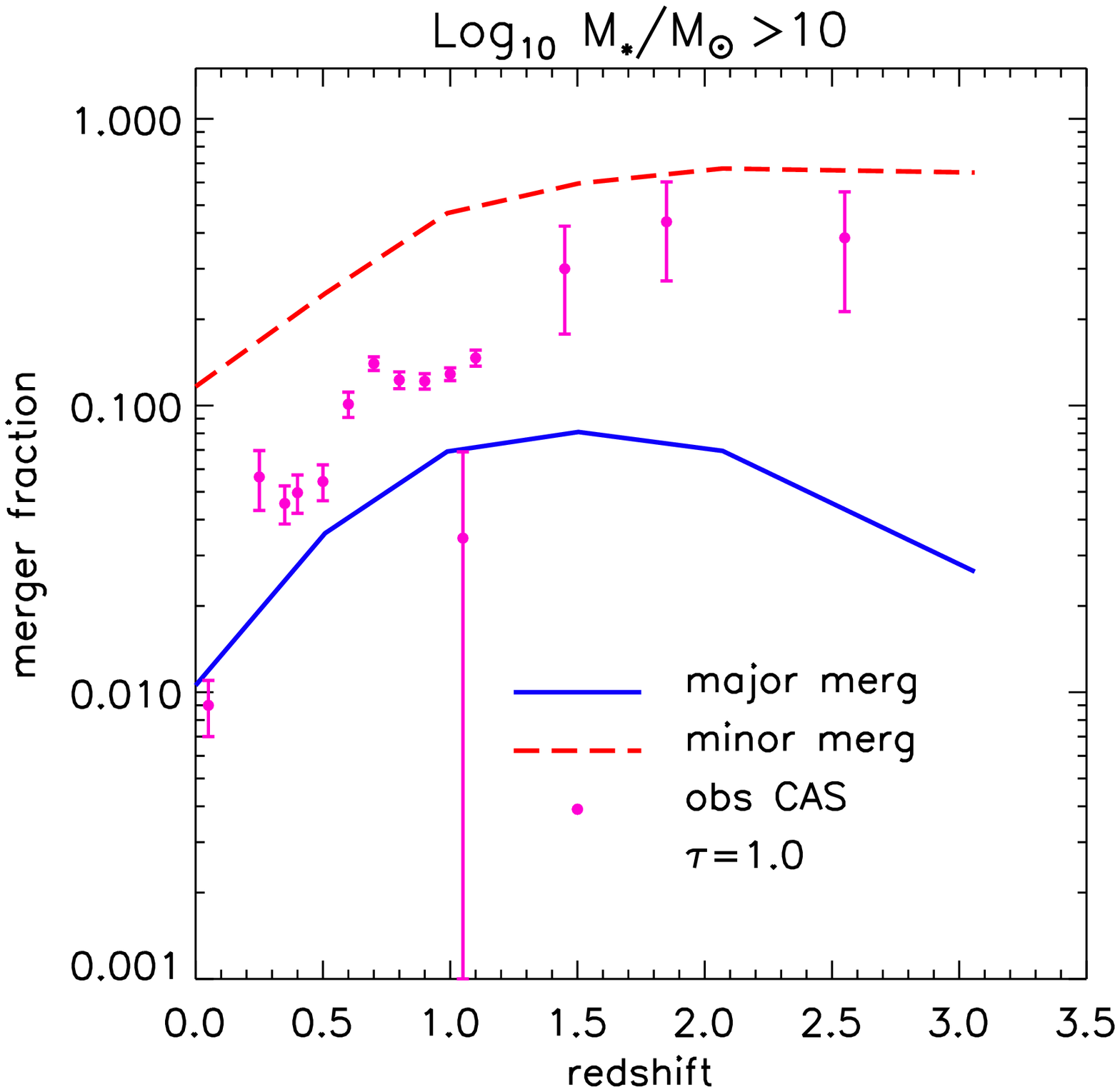}
\includegraphics[width=0.43\textwidth]{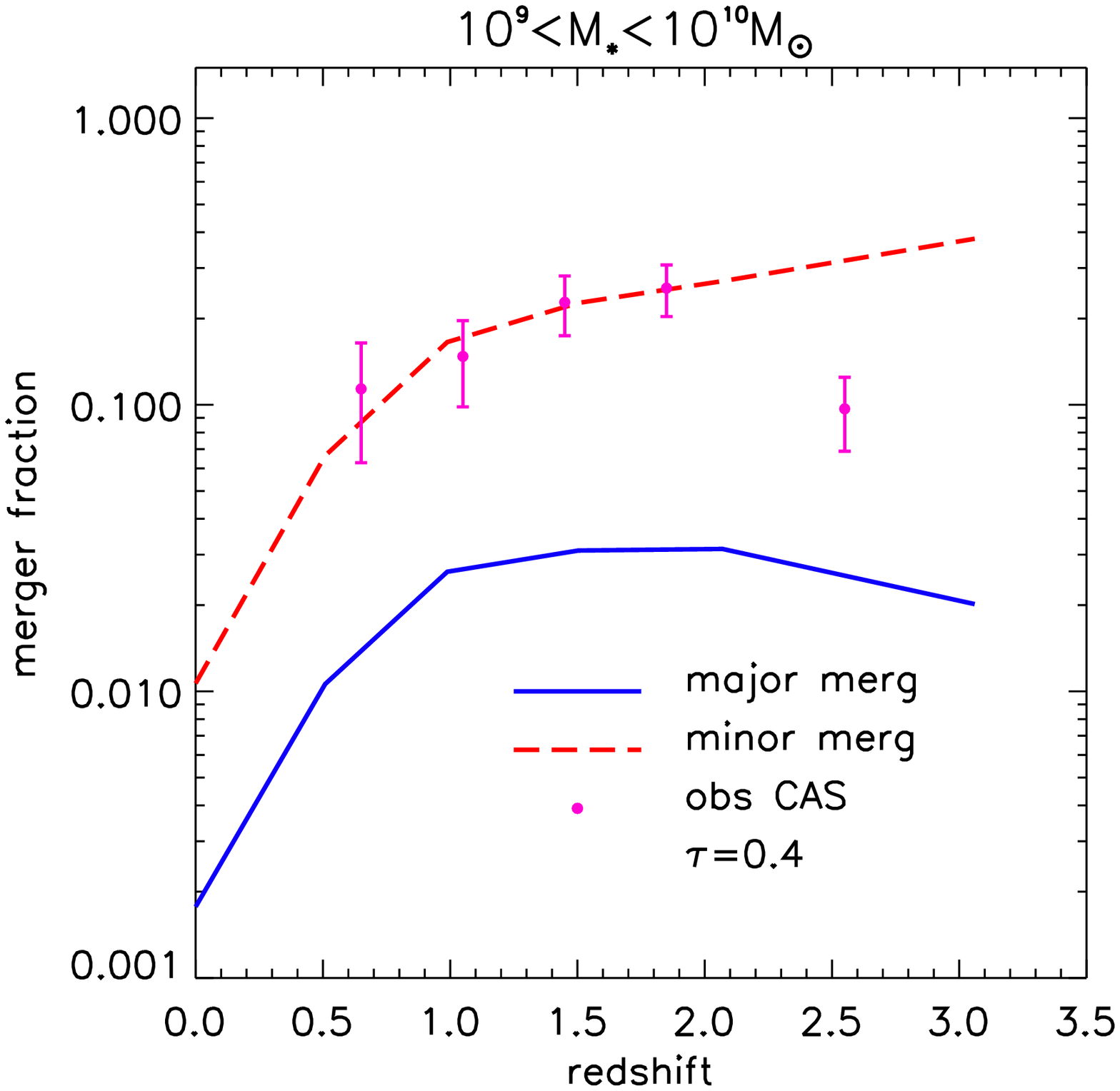}
\includegraphics[width=0.43\textwidth]{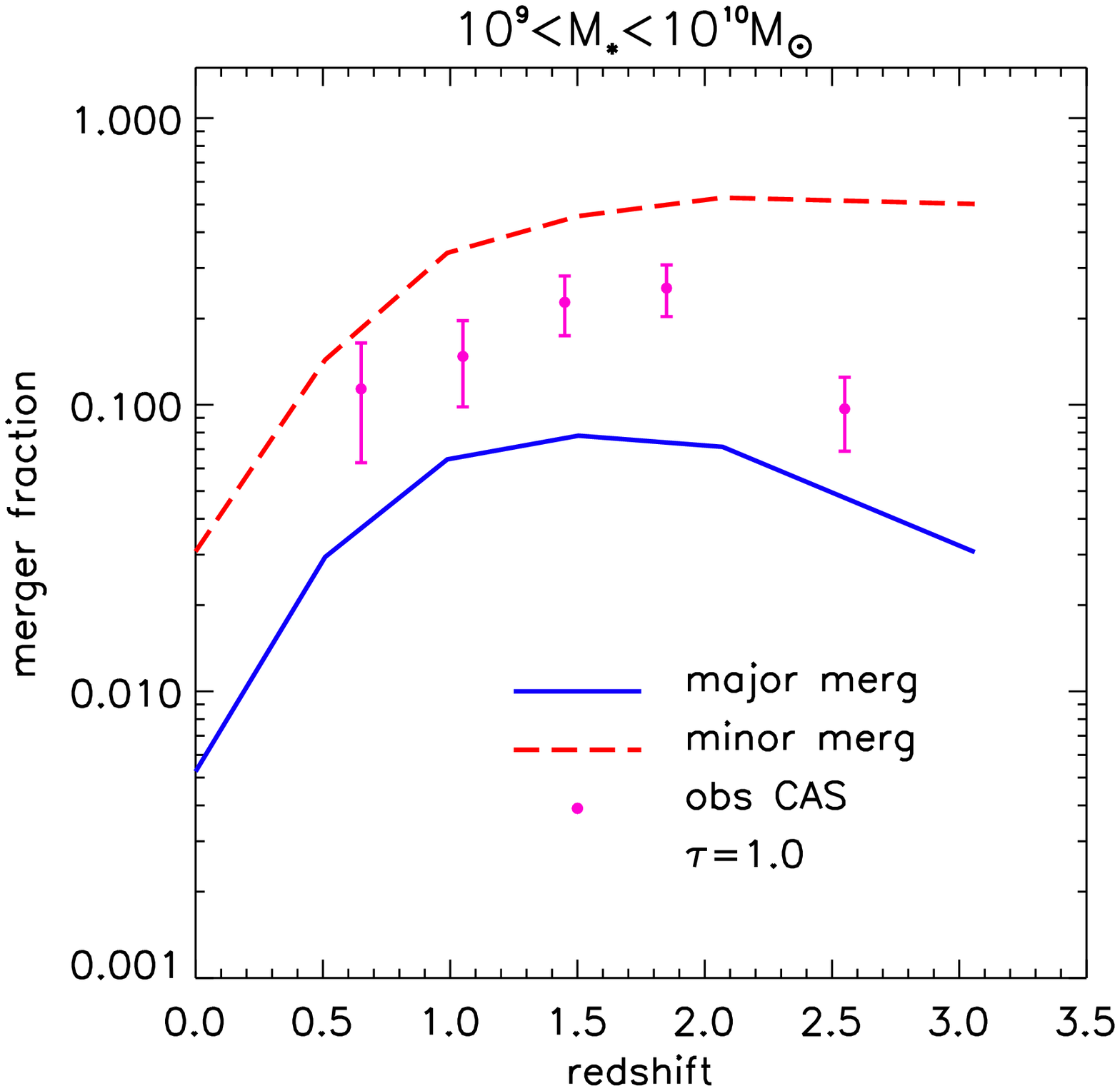}
\caption{Comparison between the observed and simulated merger fractions for galaxies with $M_{\star} > 10^{11}$ \solm\ (upper panels), $M_{\star} > 10^{10}$ \solm\ (middle panels) and $10^{9}$ \solm\ $<M_{\star}< 10^{10}$ \solm (lower panels). The dashed and solid lines indicate the minor and major merger fractions respectively. Predictions are shown in the left panels for a time-scale of $\tau_{\rm m} = 0.4$ Gyr and in the right panels for $\tau_{\rm m} = 1$ Gyr. The observed data points, described in Section \ref{data}, are filled circles for CAS results and diamonds for pair fraction results.}
\label{figcomp}
\end{figure*}

\begin{figure*}
\includegraphics[width=8.4cm]{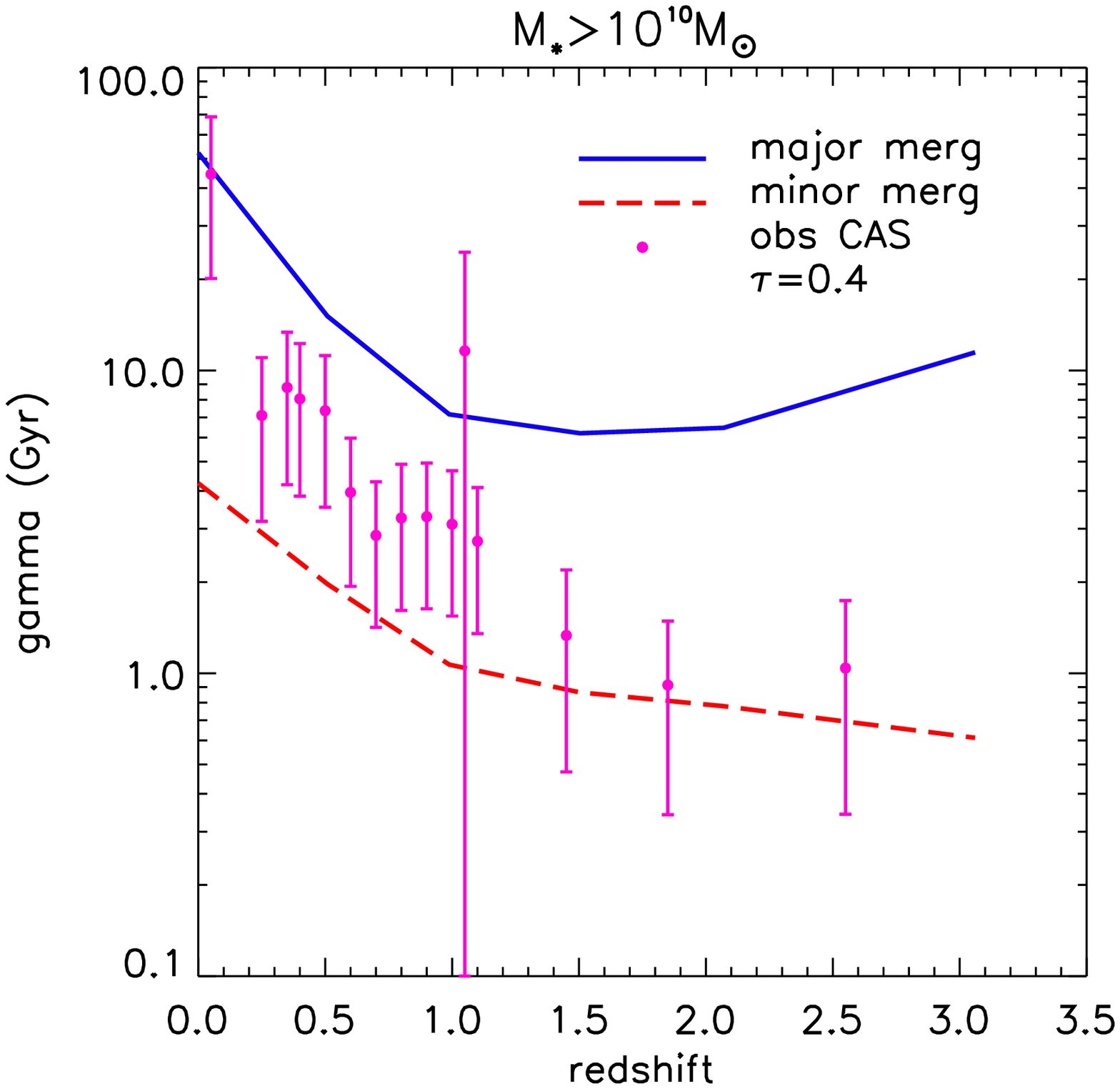}
\includegraphics[width=8.4cm]{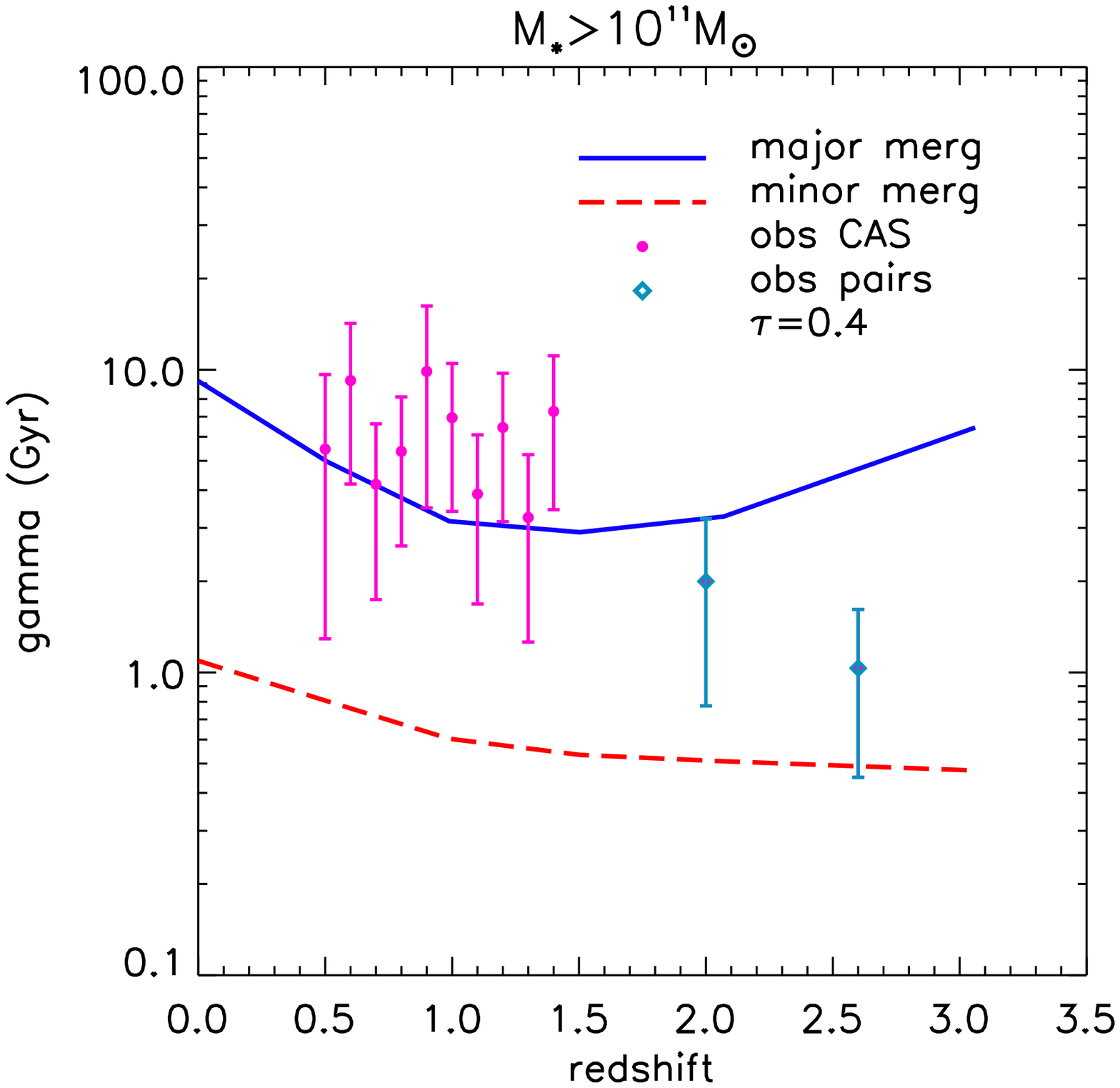}
\caption{Comparison between the observed values of the inverse of the merger rate per galaxy $\Gamma$ and predictions from the Millennium simulation as a function of redshift. The value of $\Gamma$ is roughly the time-scale between mergers. The predicted value of $\Gamma$ is calculated assuming $\tau_{\rm m} = 0.4$ Gyr. Results are shown for two mass bins: $M_{\star} > 10^{10}$ \solm\ (left panel) and $M_{\star} > 10^{11}$ \solm\ (right panel).}
\label{gamma1}
\end{figure*}

In this Subsection, we examine in detail how the observed merger fractions compare with those predicted by the Millennium simulation. 
The merger fraction, as defined in Subsection \ref{sec21}, is the fraction of galaxies undergoing a merger within a given stellar mass and redshift range. 

Fig. \ref{figcomp} shows a comparison between the observed and simulated merger fractions for galaxies with $M_{\star} > 10^{11}$ \solm\ (upper panels), $M_{\star} > 10^{10}$ \solm\ (middle panels) and $10^{9}$ \solm\ $<M_{\star}< 10^{10}$ \solm (lower panels) as a function of redshift.
To demonstrate the effect of the time-scale for merging, the simulation results in the left panels are for a time-scale $\tau_{\rm m} = 0.4$ Gyr, those 
in the right panels for $\tau_{\rm m} = 1$ Gyr.

The predicted major merger fractions for the most massive systems with $M_{\star} > 10^{11}$ \solm, shown in the upper panels of Fig. \ref{figcomp}, display a good agreement between observations and simulations at $z<2$. There is, however, a large difference between the models and the data at $z > 2$, where the observed merger fraction is about 5 to 10 times higher than predicted by 
the model.

The situation is rather different for systems with $M_{\star} > 10^{10}$ \solm, for which the predicted major merger fractions are systematically lower than the observed values by a relatively large factor, independently of the time-scale considered. The predicted major merger fraction for a time-scale $\tau_{\rm m} = 0.4$ Gyr is about an order of magnitude smaller than observed at $z\sim 0$ and the difference increases significantly at higher redshifts. 
It is most likely a coincidence that in this mass range the observed merger fractions nearly match the predicted minor merger fractions.
The minor merger fraction changes slope at $z \sim 1$, with a fast decline with decreasing redshift at $z<1$ and a slow increase with redshift at $z>1$.

As already mentioned, the largest uncertainty in estimating merger fractions in simulations is the uncertainty in the time-scale $\tau_{\rm m}$ to which
the observations are sensitive.
Fig. \ref{figcomp} shows that the observed major merger fractions are a factor of a few higher than in the Millennium even for a time-scale of $\tau_{\rm m} = 1.0$ Gyr. This is an indication that in this mass range either there are not enough major mergers in the simulations, or a larger number of minor mergers are counted as major in the observations.    Alternatively, if the time-scales
we assume are incorrect, they would have to be $\tau_{\rm m} = 4-10$ Gyr to
match the predictions.

Finally, in the bottom panels of Fig. \ref{figcomp}, we show the merger fraction evolution for galaxies with $10^{9}$ \solm\ $<M_{\star}< 10^{10}$ \solm. As we mentioned earlier, the simulated galaxies with stellar masses lower than about $10^{9.5}$ \solm\ are affected by resolution and their merger history might not be fully resolved. Consequently, the model results shown here most likely underestimate the real merger fractions. With this in mind, the observed major merger fractions are about an order of magnitude higher than the predicted ones, confirming what is seen for the mass range $M_{\star}>10^{10}$ \solm. However, it is worth pointing out that the overall merger fraction evolution for systems in both mass ranges has a similar shape as the observed one. We discuss this further in the next Section.

The merger fractions at $z>2$, that is the two data points at the highest redshifts in all ranges, are taken from \citet{bluck2009}, who measure merger fractions from galaxy pairs in the GOODS NICMOS Survey.
The estimate of the merger fraction with galaxy pairs is somewhat less likely to be biased by misinterpreting the nature of the merger than with morphological methods. However, other uncertainties might affect the estimate of merger fractions using pairs. \citet{manfred2008} argue that the time-scale for merging generally considered by observers is underestimated by at least a factor of two. This means that the two high redshift data points might be overestimating the merger fractions by a similar factor.
Since the difference between the observed and predicted merger fractions at $z=2.5$ is about a factor of ten, it is likely that there is an additional discrepancy that remains unexplained even after taking into account the correction suggested by \citet{manfred2008}.
Changing the merging time-scale from 0.4 Gyr to 1.0 Gyr, as in Fig. \ref{figcomp}, does not help solve the discrepancy, as the predicted merger fractions are then too low at high redshift and slightly too high at low redshifts.

% merger rate per galaxy

A few studies (\citealt{cons2008}; \citealt{bluck2009}) have recently investigated the behaviour of the inverse of the merger rate per galaxy, defined as $\Gamma = \tau_{\rm m}/f_{\rm gm}$.
The merger rate per galaxy is equivalent to the merger rate without the dependence on the stellar mass function and is basically the average number of mergers a galaxy experiences per unit time. As such, $\Gamma$ represents the average amount of time between mergers, and is independent of the merging time-scale $\tau_{\rm m}$ in the models. We show results for $\Gamma$ in Fig. \ref{gamma1} for the stellar mass ranges $M_{\star} > 10^{10}$ \solm\ and $M_{\star} > 10^{11}$ \solm.

In all the stellar mass bins we consider in the models, $\Gamma$ is high at high redshift, declines to a minimum at around $z \sim 1.5$ and then rises again at lower redshifts for major mergers. In other words, the predicted time between mergers at $z>2$ is longer than it is at $z \sim 1.5$.  The predicted
$\Gamma$ steadily increases with decreasing redshift at $z<1.5$, for all mass ranges and for both major and minor mergers.
On the other hand, at $z>1.5$, the predicted $\Gamma$ slowly increases for major mergers, and decreases for minor mergers. This is consistent with observations of massive galaxies (e.g., \citealt{cons2008}).
The predicted $\Gamma$ for major mergers is about ten times larger than observed for galaxies with $M_{\star} > 10^{10}$ \solm, while it matches the observations well at low redshifts for galaxies with $M_{\star} > 10^{11}$ \solm\ at $z < 2$.
This is a further indication that the merging history of galaxies with $M_{\star} < 10^{11}$ \solm\ do not agree in simulations and observations.

\subsection{Simulated Merger Fraction Parameterisations}
\label{param}

A common thing to try when examining the evolution of
the merger fraction is to parameterise the evolution in various
ways. As discussed in \citet{cons2008}, the two
most popular ways to parameterise the merger fraction are through a
power-law and through a combined exponential/power-law 
Press-Schechter-like function. 
The power-law fit is by far the most common method,
but is becoming less fashionable as it appears that the observed merger
fraction turns over at high redshifts \citep{cons2008},
while a power-law continues to increase at all redshifts.
The form of the power-law evolution is given by:

\begin{equation}
f_{\rm m}(z) = f_{0} \times (1+z)^{m}
\end{equation}
where $f_{\rm m}(z)$ is the merger fraction at a given
redshift, $f_{0}$ is the merger fraction at $z = 0$ and
$m$ is the power-law index for characterising the merger
fraction evolution. Investigating how to parameterise the
increase in the predicted merger fraction evolution is another way to
determine how the observed and predicted merger fractions 
differ. We find from previous work that the index
$m$ on this power-law increase in the merger fraction is
typically $m \sim 2-3$, which has been found using a variety of techniques (e.g., \citealt{cons2008}; \citealt{cons2009} and references therein). 

We find that the merger fraction in the Millennium simulation can be characterised by a power-law index $m = 1.6$ for $z\leq 1$ and $m=0.99$ at
$1<z\leq 3$ for galaxies with stellar masses $M_{\star} > 10^{11}$ \solm, 
and $m = 3$ for galaxies with masses $M_{\star} > 10^{10}$ \solm\ at $z < 1$.
These power law fits are only valid up to the merger fraction
turnover at $z \sim 1 - 1.5$.
While the amplitude of the observed merger fractions are higher than the simulation, the slope is fairly similar to the predictions.

Another way of characterising the merger fraction evolution, which
is based on theoretical arguments that use the Press-Schechter formalism
\citep{carlberg1990}, is a combined power-law exponential evolution of the form:
\begin{equation}
f_{\rm m} = \alpha \left( 1+z \right) ^{m} \times {\rm exp}\left[ \beta\left( 1+z \right) ^{2}\right].
\end{equation}
This combined power-law/exponential description reproduces the observations better than a simple power-law \citep{cons2006} and fits the merger fraction predictions, as well as the data (e.g., \citealt{cons2009}).
The behaviour of the merger fraction evolution can be interpreted as either the result of mergers occurring later for massive galaxies in haloes, due to the dynamical friction time-scales, or to the fact that there are not many
very massive galaxies to merge with at early times, resulting in
a lowered merger fraction. In general, it appears that all galaxies, with
the exception of the most massive ones with $M_{\star} > 10^{11}$ \solm, have
a turnover in their merger fraction history and can be fit by 
an exponential/power-law, which is likely the correct form for parameterising the merger fraction history.

\subsection{Merger Rate Evolution}
\label{rate_evolution}

\begin{figure*}
\includegraphics[width=8.4cm]{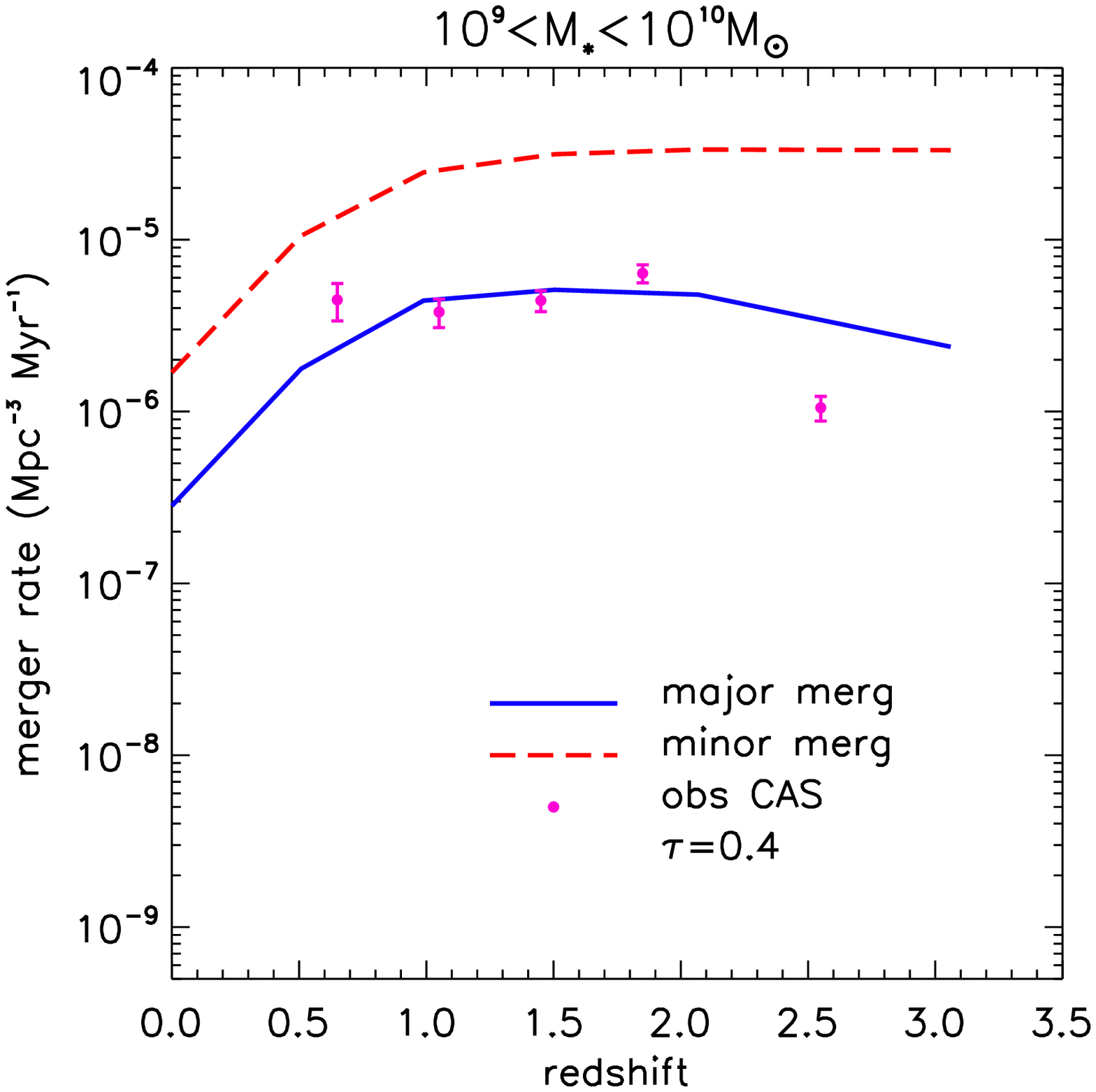}
\includegraphics[width=8.4cm]{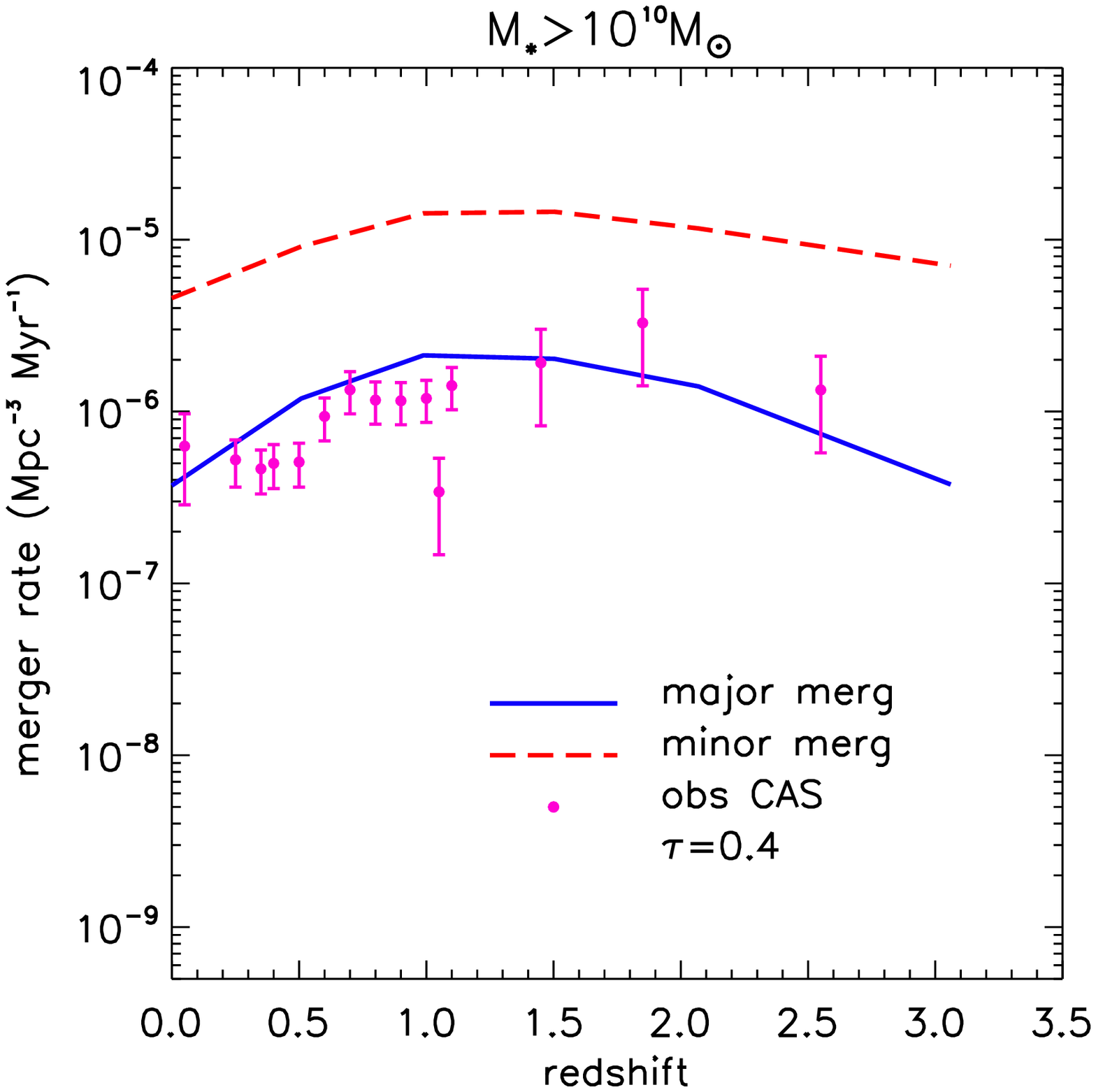}
\includegraphics[width=8.4cm]{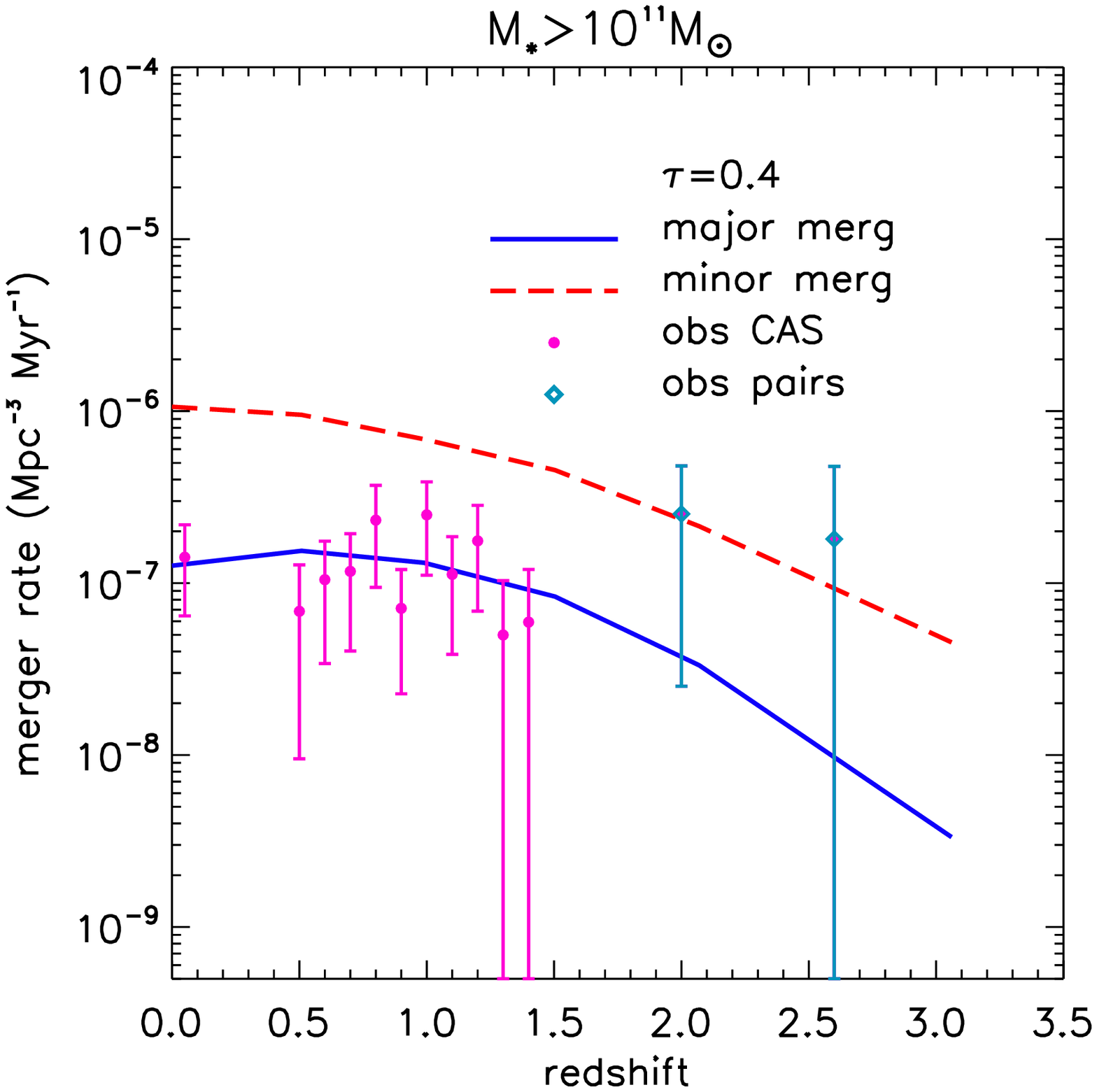}
\caption{Comparison between the observed merger rates in units of co-moving Mpc$^{3}$ per Myr and predictions from the Millennium simulation. The different lines are the same as in previous figures and show the major merger rate (solid line) and the minor merger rate (dashed line) in three different mass ranges: $10^{9}$ \solm\ $<M_{\star}< 10^{10}$ \solm\ (upper left panel), $M_{\star}> 10^{10}$ \solm\ (upper right panel) and $M_{\star}> 10^{11}$ \solm\ (lower panel).}
\label{lastfig}
\end{figure*}

In this Subsection we present results for the merger rate in the simulations and compare them to the observational data.

The merger rate \rate\ is currently the most uncertain merger quantity to measure in observations. Estimates will likely become more straight-forward as our understanding and knowledge of galaxy number densities, merger fractions and merger time-scales improve.
On the other hand, the merger rate is a relatively easy quantity to measure in simulations, once a proper definition has been agreed upon.
Since \rate\ is calculated by dividing the galaxy merger fraction, which scales with the time-scale, by the time-scale, the merger rate is independent of the time-scale $\tau_{\rm m}$ used to measure the merger fractions.
This is true under the assumption that the merger rate slowly varies with time and can be considered approximately constant within a short time-scale. However, the assumption breaks down for large time-scales, because the merger rate itself is not intrinsically constant over large time spans.

The comparison of the predicted and observed merger rates is shown in Fig. \ref{lastfig}. The predicted merger rates for the $10^{9}$ \solm\ $<M_{\star}< 10^{10}$ \solm\ galaxies (upper left panel) agree well with the data, despite the shortcomings due to lack of resolution in the simulation. The results for the other mass bins, that is $M_{\star}> 10^{10}$ \solm\ (upper right panel) and $M_{\star}> 10^{11}$ \solm\ (lower panel), agree with the observations within the error bars.

The merger rates display a surprisingly good agreement between the observations and the Millennium simulation, which is not seen for the merger fractions of galaxies with $M_{\star}< 10^{11}$ \solm.
Since the observed and predicted merger fractions agree only for the most massive galaxies, the agreement for the lower mass bins is likely  
coincidental.
As discussed in Subsection \ref{definition}, the merger rate is a function of the galaxy merger fraction $f_{\rm gm}$, the number density of galaxies $n_{\rm gm}$ and the time-scale for merging $\tau_{\rm m}$. Any one of these quantities can affect the measurement of the predicted and observed merger rates.
The merging time-scale is constant in our model, but we cannot exclude an inconsistency with the time-scale assumed by the CAS and pair methods. However, this reason alone is unlikely to explain the discrepancy between the merger fractions and rates in the lowest mass bins, given the agreement at the highest masses.

\begin{figure}
\includegraphics[width=8.4cm]{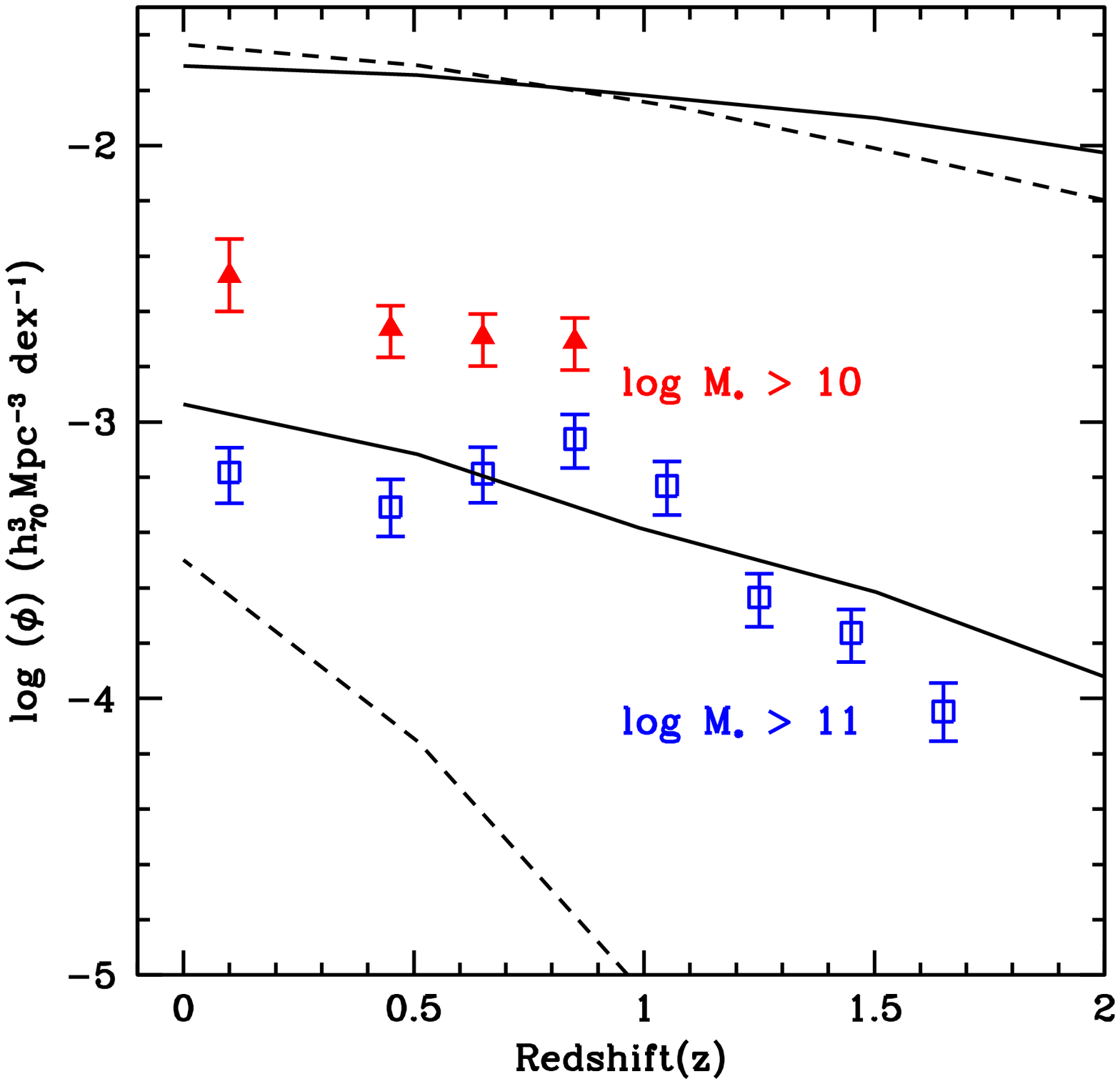}
\caption{The observed stellar mass density as a function of redshift for galaxies with $M_{\star}>10^{10}$ \solm\ (filled triangles) and $M_{\star}>10^{11}$ \solm\ (empty squares). Data points are from \citet{cons2007}. The solid line shows results for the model of \citet{bertone2007}, the dashed line for that of \citet{delucia2007}.}
\label{massf}
\end{figure}

The one quantity we have not examined yet is the predicted number density of galaxies within the Millennium simulation. Since we are selecting galaxies through a stellar mass cut, the predicted number density of these systems affects the final values of the merger rates.
We examine this in Fig. \ref{massf}, which shows a comparison between the
Millennium predictions and observations for the number density of galaxies with $M_{\star}>10^{10}$ \solm\ (filled triangles) and $M_{\star}>10^{11}$ \solm\ (empty squares). Data points are taken from \citet{cons2007}. The solid line shows results for the model of \citet{bertone2007}, the dashed line for that of \citet{delucia2007}.

The model of \citet{bertone2007} roughly predicts the number density of galaxies with $M_{\star}>10^{11}$ \solm, while the model of \citet{delucia2007} underpredicts it at all redshifts. Both models overestimate the observed values for galaxies with $M_{\star}>10^{10}$ \solm\ by almost an order of magnitude.
This results in the high value derived for the merger rate of galaxies with $M_{\star}<10^{11}$ \solm, since, as we know from Fig. \ref{figcomp}, the predicted merger fraction is low compared to the data. We therefore conclude that the apparent agreement seen in Fig. \ref{lastfig} is partly a coincidence for the stellar mass interval $M_{\star}>10^{10}$ \solm. We believe that the agreement is instead genuine for the more massive galaxies with $M_{\star}>10^{11}$ \solm.

\section{Discussion}
\label{discussion}

Tracing the galaxy merger history observationally is just now becoming
a mature enough area of astronomy that it can be compared with theory.
The modern concept for how galaxies and structures form in the universe is based on the idea that galaxies accrete a significant fraction of their mass through mergers.
While the $\Lambda$CDM paradigm predicts many features of the observed galaxy
population successfully, many galaxy properties are difficult to explain and reproduce in models. These include the mass profiles of galaxies, the missing satellite problem \citep{moore1999} and the fact that at high redshift there are more massive galaxies than predicted (e.g., \citealt{cons2007}).
These differences might be due to the way that the physics of star formation is implemented in the models, or to the underlying dark matter halo history, or perhaps even the underlying cosmology.
Comparing the merger histories predicted by models of galaxy formation with available data is a new way to help test our understanding of galaxy formation.

As discussed in Subsection \ref{rate_evolution}, we find that major merger rate predictions match the observations fairly well. The merger fraction predictions discussed in Subsection \ref{fracs} match the observations well for galaxies with $M_{\star}>10^{11}$ \solm\ at $z<2$, but diverge significantly when less massive galaxies are considered.
For these galaxies, the observed major merger history is more substantial than predicted by the semi-analytic models of \citet{bertone2007} and \citet{delucia2007}. Previous investigations have highlighted a similar problem when examining galaxies in pairs compared with the Millennium simulation \citep{mateus2008}.

In the following we discuss several possible explanations for the discrepancy between the observed and the predicted merger rates and fractions of galaxies with intermediate masses, including observational biases in finding mergers, problems with the underlying mass assembly history and the predicted star formation history of galaxies.

\subsection{Variations due to the physical modelling of star formation and feedback}
\label{varymodel}

\begin{figure*}
\includegraphics[width=0.49\textwidth]{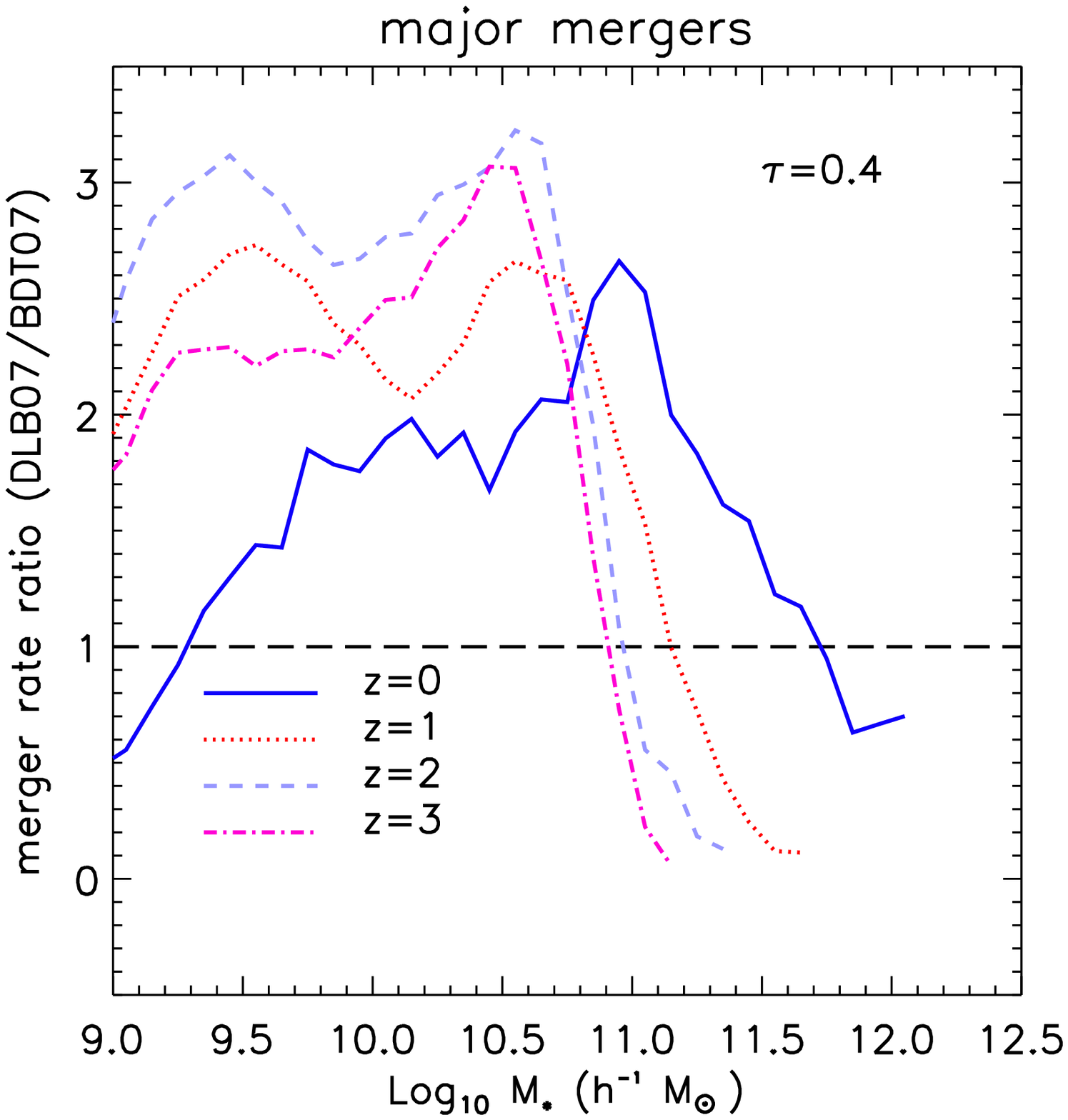}
\includegraphics[width=0.49\textwidth]{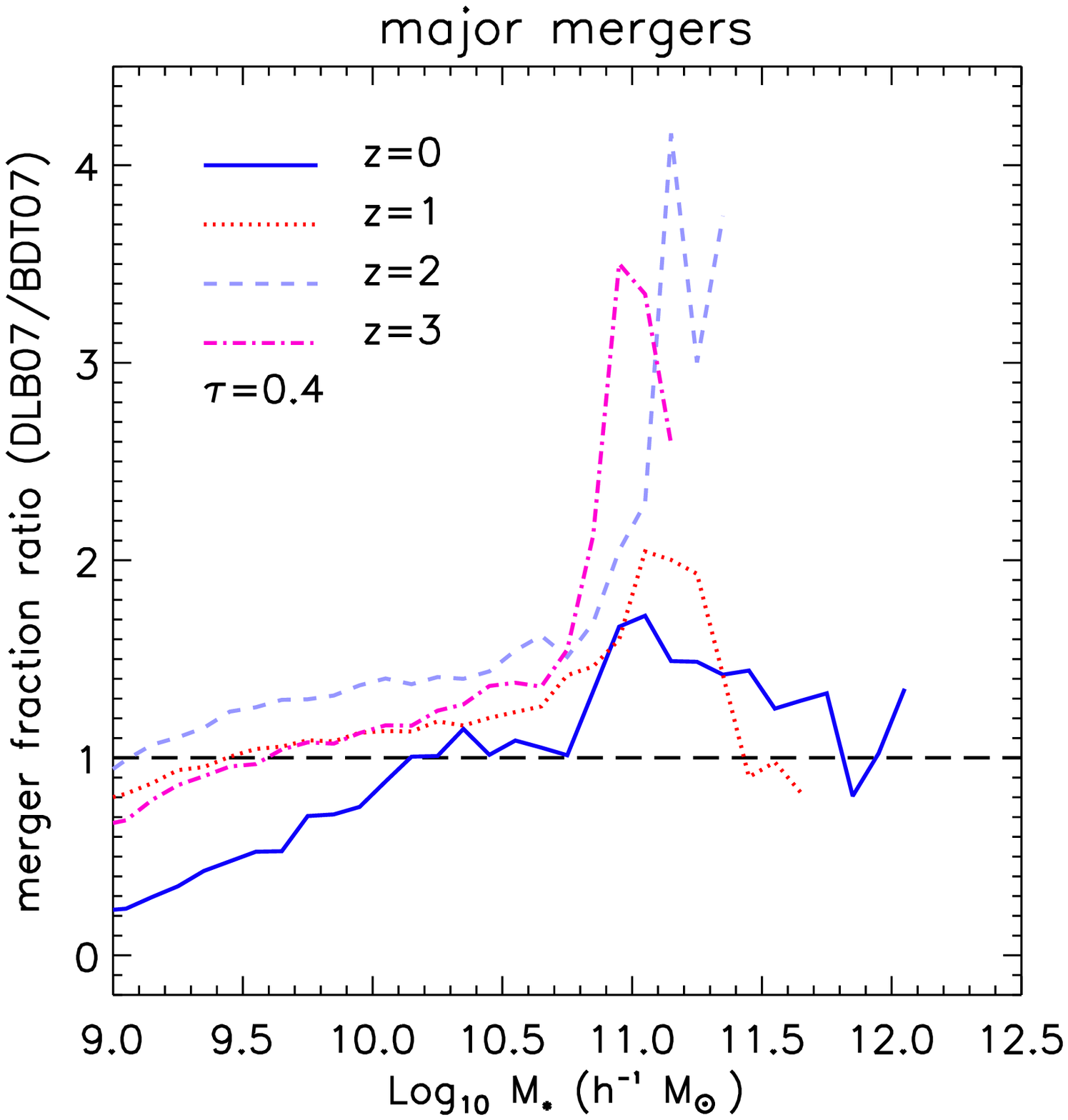}
\caption{Comparison between the major merger rates (left panel) and major merger fractions (right panel) predicted by the BDT07 \citep{bertone2007} and by the DLB07 \citep{delucia2007} models as a function of galaxy stellar mass. Results are shown for redshifts $z=0, 1, 2$ and 3.
In both panels, the $y-$ axis shows the ratio of the values predicted by the two models. The dashed horizontal line indicates where the model predictions are equal. Both figures demonstrate that in general the \citet{delucia2007} model predicts higher values for the merger rates and fractions for $M_{\star}<10^{11}-10^{11.5}$ \solm\ at almost all redshifts. Conversely, the \citet{bertone2007} model provides higher merger rates and fractions in the highest stellar mass bins.}
\label{d1d3}
\end{figure*}

One major issue we have not addressed yet is the fact that we are selecting
galaxies by their stellar mass. The merger fraction and rate vary as
a function of stellar mass, both within observed galaxies (e.g., \citealt{cons2003a}; \citealt{cons2008}) and within models (see Fig. \ref{fig2}). An uncertainty in the determination of the stellar mass of galaxies could translate in to an uncertainty in the measured merger values.
Uncertainties in the star formation history would place galaxies into too high or too low stellar mass bins and affect the mass ratio of the merger event. This in turn affects the predicted major merger fractions and rates.

In the semi-analytic model, galaxies that fall into a larger halo cannot form many stars after they become satellites, as they cannot accrete gas from their environment and quickly consume or expel the interstellar medium they might have possessed when they were central galaxies themselves. This is a well-known 
problem within semi-analytic models and is one of the causes of the excess red population of low mass galaxies on the colour-magnitude diagram, which are not seen in observations \citep{bertone2007}. As a consequence, a larger fraction of predicted merging galaxies might have lower mass ratios than what is necessary to be counted as major mergers, and a larger number of major mergers could end up being counted as minor mergers. A factor that could counteract this effect, though, is stellar mass loss through tidal stripping in cluster galaxies (\citealt{bullock2001}; \citealt{murante2007}), which is currently not accounted for in the semi-analytic model.

As mentioned in \S \ref{millennium}, the merger fractions and rates in the model of \citet{bertone2007} may differ from those of other models based on the Millennium simulation, including for example the model of \citet{delucia2007}. 
These two models differ only in the treatment of supernova feedback, but otherwise use the same merger trees and the same treatment of gas physics. Star formation, AGN feedback, gas cooling, merging and other relevant processes are done in the same way, using the same parameters. 

The SN feedback scheme is different in the following ways. In the \citet{delucia2007} model the SN feedback is based on simple formulae of energetic balance, as in other semi-analytical models. In the \citet{bertone2007} model, the evolution of galactic winds is followed by solving the equations of motion of the outflows, which are modelled as pressure-driven cosmological blastwaves when they emerge from galaxies and become momentum-driven snowploughs when cooling sets in. The treatment of SN feedback in the \citet{bertone2007} model implements a faster recycling of gas than in the \citet{delucia2007} model, which suppresses star formation more efficiently in dwarf galaxies and less efficiently in massive galaxies than the feedback prescriptions in the \citet{delucia2007} model. This results in a larger number of massive galaxies and in a shallower slope at the faint end of the stellar mass function in the \citet{bertone2007} model than in the \citet{delucia2007} model.

In Fig. \ref{d1d3} we quantify the difference between these two models by 
showing the ratio of the predicted major merger rates (left panel) and fractions (right panel) of the \citet{bertone2007} and \citet{delucia2007} models. Results are shown as a function of stellar mass for the redshifts $z=0, 1, 2$ and 3. The dashed horizontal line indicates where the model predictions are equal. The two modelled merger rate predictions differ by no more than a factor of three, in the most extreme case, with the largest difference being in the values of the merger rates at $z>0$ for galaxies with $M_{\star}<10^{11}$ \solm. As 
discussed above, a possible explanation for this difference is the higher 
abundance of galaxies in this stellar mass range in the \citet{delucia2007} model. The predicted merger fractions are roughly similar, with the important exception of a narrow stellar mass bin around $M_{\star} \sim 10^{11}$ \solm.
In the \citet{delucia2007} model the merger fractions for the $M_{\star} > 10^{11}$ \solm\ galaxies nearly match the observations if the time-scale for merging is $\tau_{\rm m} = 1$ Gyr.
Assuming $\tau_{\rm m} = 0.4$  produces results that slightly underpredict the data. The predicted major merger fractions in the \citet{delucia2007} model differ by a very minor amount from the \citet{bertone2007} model for galaxies at $M_{\star} < 10^{10}$ \solm. As a consequence, the different feedback prescriptions used in the two models do not significantly help to better match the observations at low stellar masses, where the discrepancy is largest.

\subsection{Variations in the Merger Histories}

The discrepancy between the observed and predicted major merger fractions could also result from a lack of merging events in the simulation itself, or to systematic differences in the calculations of merger fractions, or in the definition of what is a merger.

One difference between the model and the observations that we are aware of, 
and which we discussed in Section \ref{millennium}, is the definition of what 
is a major merger. Observations define mergers as 
major when the mass ratio between the merging galaxy and the central 
galaxy is greater than 1:4. In the Millennium simulation, major mergers 
happen when the mass ratio is greater than 0.3. This difference may account for some of the discrepancy in the estimates of the merger fractions and rates: if 
a mass ratio 0.25 were used also for the simulations, the estimate of the 
merger fractions would increase by about 20 per cent, as it 
would also include mergers that are currently not counted as major.
This would improve somewhat the agreement between the observed and predicted 
merger fractions in some stellar mass bins, but it would not significantly improve the overall agreement and it would not solve the discrepancy at $z>2$.
\citet{mateus2008} also points out that the observed merger rates and fractions are better reproduced by the models when mergers with a mass ratio smaller than 1:4 are included in the estimation of the merger rate.
Indeed, \citet{mateus2008} best match with the observations is achieved for a mass ratio of 1:10.

It is also not entirely clear what is the CAS major merger ratio sensitivity. This is generally assumed to be 0.25, but it could equally be 0.3, or lower. We know this is unlikely, based on comparing CAS and pairs for nearby galaxy mergers (\citealt{hernandez2005}; \citealt{depropris2007}). For a basic agreement at high redshift or at low stellar masses, the CAS method would have to be sensitive to mergers with mass ratios down to 1:10.
This is again unlikely to be the major reason for the difference between models and observations, because the variations involved are too small in comparison with the discrepancies in the results.
However, it is a further indication that the mass ratio threshold to identify major mergers is indeed a rather volatile factor both in the observations and in the simulations.

Finally, the mismatch between the observations and simulation results might be due to the inability of the Millennium simulation to produce enough major mergers. This can happen, for example, if there are too few galaxies in haloes, or if the predicted merging time-scale is too long.
The lack of galaxies in haloes can be tested by measuring the bias of massive galaxies in the Millennium simulation, which have to be more clustered than the dark matter. \citet{zehavi2005} found that the predictions of semi-analytical models agree fairly well with the clustering properties of nearby galaxies. However, \citet{coil2008} found that at high redshift simulated blue galaxies are much less clustered than real blue galaxies, especially at small scales.
As the vast majority of merging galaxies at $z > 1$ are blue \citep{cons2003b}, this is a potential problem of the models, which translates in to an underestimate of the merger history.
In fact, a lack of a strong bias, or equivalently a lower halo occupation number, implies that massive systems do not experience as many merger events as they would if they were more clustered.

\subsection{Comparison with Previous Work}
\label{comp}

In this Subsection we briefly compare our findings with results from previous works.

\citet{guo2008} examine how galaxies grow within the Millennium model. They find that the most massive galaxies have a formation history dominated by major mergers, while for less massive galaxies the same is true only at $z>1$.
\citet{manfred2008} use the model of \citet{delucia2007} to calibrate the relationship between the abundance of close galaxy pairs and the rate of galaxy mergers at high redshifts. They find that close galaxy pairs merge within a few Gyr, and that this merging time-scale only weakly increases with decreasing redshift. \citet{manfred2008} also argue that while the use of close pairs for estimating the merger history of galaxies is a reliable tool, the merging times used to calculate the merger rates in observations are at least a factor of two too short with respect to their findings, which translates in to an overestimate of the merger rates by a similar factor. \citet{patton2008} show that this is not the case when an appropriate correction factor is included in the equations.
Similar conclusions are reached by \citet{wetzel2008b}, who find that only pairs at very small separations can be considered a reliable proxy for the global merger population.

Using galaxy pair fractions, \citet{mateus2008} finds a similar qualitative 
behaviour with redshift for the merger rate as we do. \citet{mateus2008} also compares the results from the model of \citet{delucia2007} with those from the model of \citet{bower2006} and finds that the \citet{bower2006} model better reproduces the qualitative behaviour of the merger rate with redshift than the 
\citet{delucia2007} model, although the predicted values are still lower than observed. This is interesting, because \citet{bower2006} use different merger trees than \citet{delucia2007} and \citet{bertone2007} and have different prescriptions for merging that might better reproduce the evolution of the observed galaxy population.

Other studies have compared the observed galaxy merger properties with the Millennium model (\citealt{genel2008}; \citealt{patton2008}).
\citet{patton2008} compare the number of close galaxy pairs (with separations smaller than 20 \hm\ kpc) in the simulation and in a sample of SDSS galaxies at low redshift and find good agreement. \citet{patton2008} estimate that at least 90 per cent of major mergers occur between galaxies fainter than $L_{\star}$.
\citet{genel2008} find that there are not enough major mergers in the Millennium simulation to transform the population of $z=2$ galaxies in to the elliptical galaxies observed at $z=0$, which implies that in the model internal evolution must play a more important role to shape their properties than in the real universe.

\section{Conclusions}
\label{conclusion}

We have carried out a comparison between galaxy merger history predictions
based on the Millennium simulation and a collection of observed galaxy merger 
fractions and rates taken from \citet{cons2003b}, \citet{cons2008}, \citet{bluck2009} and \citet{cons2009}.
The observed merger rates and fractions are measured using the CAS system  and by measuring the number of galaxy pairs at $z>2$.

Our main results for the Millennium galaxy formation model of \citet{bertone2007} can be summarised as follows:

\begin{itemize}

\item The predicted major merger rates and fractions vary significantly with time and stellar mass. Massive galaxies experience on average more frequent major merger events than less massive galaxies, although the merger rate in general is higher for less massive galaxies.

\item The model predicts that the major merger fraction increases with redshift
between $z=0$ and about $z\sim 1.5$, reaches a peak at about $z\sim 1.5$ and then declines at higher redshifts.
This result is independent of the time-scale $\tau_{\rm m}$ over which mergers are identified and is common to all the stellar mass ranges we consider.
This is roughly consistent with the behaviour of the observed data at $z < 2$, but not at higher redshift, where the observed major merger fractions continue to increase or remain high at larger redshifts.
The predicted values for the major merger fractions match the observed ones for galaxies with $M_{\star} > 10^{11}$ \solm, but underestimate them when less massive galaxies are considered.

\item The major merger rates predicted by the Millennium simulation roughly agree with the observed rates in all stellar mass ranges and reproduces  well their evolution with redshift. However, since the major merger fractions agree with observations only for galaxies with $M_{\star} > 10^{11}$ \solm, the agreement at the lower stellar masses is likely  coincidental.

\item We discuss several issues that might introduce uncertainties in the measurement of the merger fractions in the models and in the data.
Some disagreement can be alleviated at some redshifts and stellar mass ranges, but becomes worse in other cases.
Among other things, a different implementation of SN feedback in the semi-analytic model, such as that implemented by \citet{delucia2007}, produces variations in the results by up to a factor of a few.
The mass ratio to which different techniques are sensitive might also be a source of uncertainty, but probably only a minor one.
Likewise, the stellar masses in the Millennium simulation may be affected by processes, such as strangulation and tidal stellar stripping, that are not realistically implemented, and that ultimately affect the mass ratio and the number density of mergers. It is also possible that models are not producing enough major mergers in galaxies with $M_{\star} < 10^{11}$ \solm.
\end{itemize}

The fact that the merger predictions do not match the observed major merger history is potentially an explanation for other known problems in the semi-analytic models associated with the Millennium simulation. These include, for example, the overproduction of red, low mass galaxies in colour-magnitude diagrams. This problem might be partially alleviated if the galaxy merger history is more substantial than predicted.

\section*{Acknowledgements}
We would like to thank G. De Lucia, S. Foucaud, P. Jonsson, F. Pearce, J. Primack and S. White for useful discussions and the anonymous Referee for many constructive comments that helped to improve the manuscript.
SB acknowledges support from NSF Grant AST-0507117. 
The Millennium simulation was carried out by the Virgo Consortium at the Max Planck Society in Garching. Data on the galaxy population produced by the models used in this work, as well as on the parent halo population, are publicly available at http://www.mpa-garching.mpg.de/millennium/.

\label{lastpage}

\end{document}